\documentclass[preprint,10pt]{elsarticle}

\usepackage{amssymb}

\usepackage{caption}
\usepackage{float}
\usepackage{subcaption}
\usepackage{siunitx}
\usepackage{amsmath}
\usepackage{xcolor}
\usepackage{multirow}
\usepackage[numbers]{natbib}
\usepackage{pbox}
\usepackage{ragged2e}
\usepackage{gensymb}
\usepackage{vmargin}
\usepackage{comment}
\usepackage{ulem}

\newcommand{\rebuttal}[1]{\textcolor{black}{#1}}

\usepackage{lineno}

\journal{Applied Mathematical Modelling} 

\begin{document}

\begin{frontmatter}



\title{Reduced-order models of \rebuttal{wall} shear stress patterns in the left atrial appendage from a data-augmented \rebuttal{atrial} database
}


\author[1]{Jorge Dueñas-Pamplona\corref{corr1}}
\cortext[corr1]{Corresponding author}
\ead{jorge.duenas.pamplona@upm.es}

\address[1]{Departamento de Ingenería Energética, Universidad Politécnica de Madrid, Madrid, Spain}
\address[2]{Departamento de Ing.  Mecánica, Energética y de los Materiales, Universidad de Extremadura, Badajoz, Spain}
\address[3]{Department of Mechanical Engineering, University of Washington, Seattle, WA 98195, US}
\address[4]{Department of Bioengineering, University of Washington, Seattle, USA}
\address[5]{Departamento de Ingeniería Energética y Fluidomecánica, Universidad de Valladolid, Valladolid, Spain}
\address[6]{Instituto de Computación Científica Avanzada de Extremadura (ICCAEX), Badajoz, Spain}
\address[7]{Department of Aerospace Engineering, University Carlos III of Madrid, Leganés, Spain}
\address[8]{Institute for Stem Cell and Regenerative Medicine, University of Washington, Seattle, USA}
\address[9]{Center for Cardiovascular Biology, University of Washington, Seattle, USA}

\author[2]{Sergio Rodríguez-Aparicio}
\author[3]{Alejandro Gonzalo}
\author[4]{Savannah F. Bifulco}
\author[5]{Francisco Castro}
\author[2,6]{Conrado Ferrera}
\author[7]{Óscar Flores}
\author[4,8,9]{Patrick M. Boyle}
\author[5]{José Sierra-Pallares}
\author[1]{Javier García García}
\author[3]{Juan C. del Álamo}

\begin{abstract}
\noindent\textbf{Background:} Atrial fibrillation (AF) is the most common sustained cardiac arrhythmia, affecting over 1\% of the population. It is usually triggered by irregular electrical impulses that cause the atria to contract irregularly and ineffectively. It increases blood stasis and the risk of thrombus formation within the left atrial appendage (LAA) and aggravates adverse atrial remodeling. Despite recent efforts, LAA flow patterns representative of AF conditions and their association with LAA stasis remain poorly characterized.  

\noindent\textbf{Aim:} To develop reduced-order data-driven models of LAA flow patterns \rebuttal{during atrial remodeling in order to uncover flow disturbances concurrent with LAA stasis that could add granularity to clinical decision criteria.}

\noindent\textbf{Methods:} We combined a geometric data augmentation process with projection of results from 180 CFD atrial simulations on \rebuttal{a} universal LAA coordinate (ULAAC) system. The projection approach enhances data visualization and facilitates direct comparison between different anatomical and functional states. ULAAC projections were used as input for a proper orthogonal decomposition (POD) algorithm to build reduced-order models of hemodynamic metrics, extracting flow characteristics associated with AF and non-AF \rebuttal{anatomies}. 

\noindent\textbf{Results:} We verified that the ULAAC system provides an adequate representation to visualize data distributions on the LAA surface and to build POD-based reduced-order models. These models revealed significant differences in LAA flow patterns \rebuttal{for atrial geometries that underwent adverse atrial remodeling and experienced elevated blood stasis.}
Together with anatomical morphing-based patient-specific data augmentation, this approach could 
\rebuttal{facilitate data-driven analyses to identify flow features associated with thrombosis risk due to atrial remodeling.} 
\end{abstract}




\end{frontmatter}

\section{Introduction} \label{sec:introduction}

Cardiovascular diseases are the leading cause of death worldwide. Of these, atrial fibrillation (AF) is the most common cardiac arrhythmia, affecting more than 35 million people worldwide \cite{benjamin2019heart}. More than 20 \% of the 18 million yearly ischemic strokes are estimated to be caused by thrombi that originate in the left atrium (LA) during an episode of AF. Furthermore, another 30 \% of ischemic strokes are suspected to originate in the atrium in patients with subclinical AF or sinus rhythm \cite{kamel2016atrial}. 

Atrial fibrillation usually begins with a dysfunction in electrical impulses, which generates a weak and irregular pattern of atrial contraction and relaxation. These episodes cause blood stasis in the left atrial appendage (LAA), promoting thrombosis. Consequently, the risk of stroke is five times higher in patients diagnosed with AF compared to those without AF \cite{Wolf1991}. When occurring over periods of months or years, AF episodes can trigger adverse atrial remodeling, a variety of tissue alterations that disrupt the biomechanical and electrical function of the myocardium. This remodeling exacerbates the arrhythmogenic substrate for AF and the hemodynamic substrate for thrombosis \cite{boyle2021fibrosis}. 

The LAA is the most frequent site of atrial thrombosis. It is a protruding cavity inside the LA, a residue of the embryonic developmental stage \cite{Al-Saady1999}. The morphology of LAA is highly variable, varying both in shape and size from patient to patient \cite{lupercio2016left}. Recent work hypothesized that LAA might play an important role in circulatory system homeostasis and hemodynamics \cite{murtaza2020role,lakkireddy2018left}. It also acts as a contractile reservoir or decompression chamber, depending on the cardiac cycle phase \cite{hoit2014left}. This effect worsens with time due to atrial remodeling, which tends to increase its volume and reduce its contractility. 

In recent years, computational fluid dynamics (CFD) models of increasing complexity have been developed to study flow patterns within the heart. These include models of the whole left heart \cite{mihalef2011patient}, models of two-cavity including LA and LV \rebuttal{\cite{Vedula2015, chnafa2014image, bucelli2022mathematical}}, and models of one-cavity that usually include LA or LV \rebuttal{\cite{chnafa2014image,Otani2016,Lantz2018a}}. Of the latter, most have focused on flow patterns in the LV \cite{seo2014effect, vedula2016effect}, although there has recently been an increasing number of studies focusing on LA hemodynamics, specifically studying stasis within the LAA \rebuttal{\cite{Otani2016, Bosi2018,Garcia-Isla2018,Masci2019, duenas2021comprehensive, Koizumi2015, corti2022impact, garcia2021demonstration, duran2023pulmonary}}. These were motivated by clinical evidence of the relationship between LAA morphology and the risk of thrombosis in patients with AF \cite{DiBiase2012, yamamoto2014complex}. These studies have focused on finding the relationship between LAA morphology and stasis \rebuttal{\cite{Garcia-Isla2018, Masci2019,garcia2021demonstration}}, proposing new metrics to quantify this stasis \rebuttal{\cite{corti2022impact, duenas2022morphing, Mill2020}}, \rebuttal{contrasting common modeling assumptions in fluid-dynamic simulations of the LA \cite{duenas2021comprehensive, duran2023pulmonary, duenas2021boundary, gonzalo2022non}, or performing sensitivity analyses \cite{mill2021sensitivity, khalili2023importance}}. Recently, multiphysics models integrating fluid-structure interaction together with electrophysiological or biomechanical mechanisms related to AF are being proposed \rebuttal{\cite{bucelli2022mathematical,corti2022impact, gonzalo2022non, feng2019analysis}}.  

The studies above have shed light on the mechanisms of AF and informed us about proper model generation, boundary conditions, and numerics to accurately reproduce LA flow dynamics. However, blood flow and thrombosis within the LAA are multifactorial processes \cite{boyle2021fibrosis}, and existing CFD studies have yet to produce a predictive understanding of their anatomical and functional determinants\rebuttal{, which at the same time are linked to the advance of adverse atrial remodeling. Another} particularly limiting factor is that the diversity of LAA morphologies makes it challenging to compare simulation results from different patients or, even for the same patient, between different cardiac conditions \cite{bifulco2021computational}. 

Considering these challenges, \rebuttal{we propose a new approach towards developing a data-driven analysis capable of capturing the flow features associated with adverse atrial remodeling.} For this purpose, we used the atrial morphing methodology proposed by Dueñas-Pamplona et al. \cite{duenas2022morphing} with a new framework that allowed us to compare hemodynamic variables between patients and cardiac conditions. Following the ideas proposed by \cite{borja2020automatic} for the left ventricle, we employ proper orthogonal decomposition (POD) to reduce flow dimensionality, identify distinct flow patterns associated with different phenotypes, and classify patients according to their LAA flow patterns. 

To facilitate this process \cite{acebes2021cartesian}, we map the LAA surface onto a unified coordinate system (Universal Left Atrial Appendage Coordinates, or ULAAC), which is an extension of the Universal Atrial Coordinate (UAC) system developed by Roney et al. \cite{roney2019universal}. \rebuttal{We illustrate this novel approach by analyzing flow fields obtained by running CFD simulations on LA anatomical models of patients with and without adverse atrial remodeling.}

\rebuttal{Current medical procedures to estimate ischemic stroke risk are based on demographic and clinical factors and do not consider patient-specific information about LA hemodynamics or the existence of atrial remodeling. This approach can offer direct insight into specific flow characteristics induced by anatomical remodeling, shedding light on the relationship between the different variables that influence stasis patterns due to remodeling, such as LAA morphology, LAA volume, ostium area, etc. Therefore, this approach has the potential to add granularity to clinical thresholds to predict the risk of stroke in patients with AF.}

The paper is organized as follows. Medical imaging, atrial model morphing, hemodynamic indices, CFD simulations, ULAAC projection, and POD are described in Section \ref{sec:methods}. The results are presented in Section \ref{sec:results} and discussed in Section \ref{sec:discussion}, reflecting the main conclusions achieved.

\section{Methods}\label{sec:methods}

\begin{figure}[t]
\centerline{\includegraphics[width=1\linewidth]{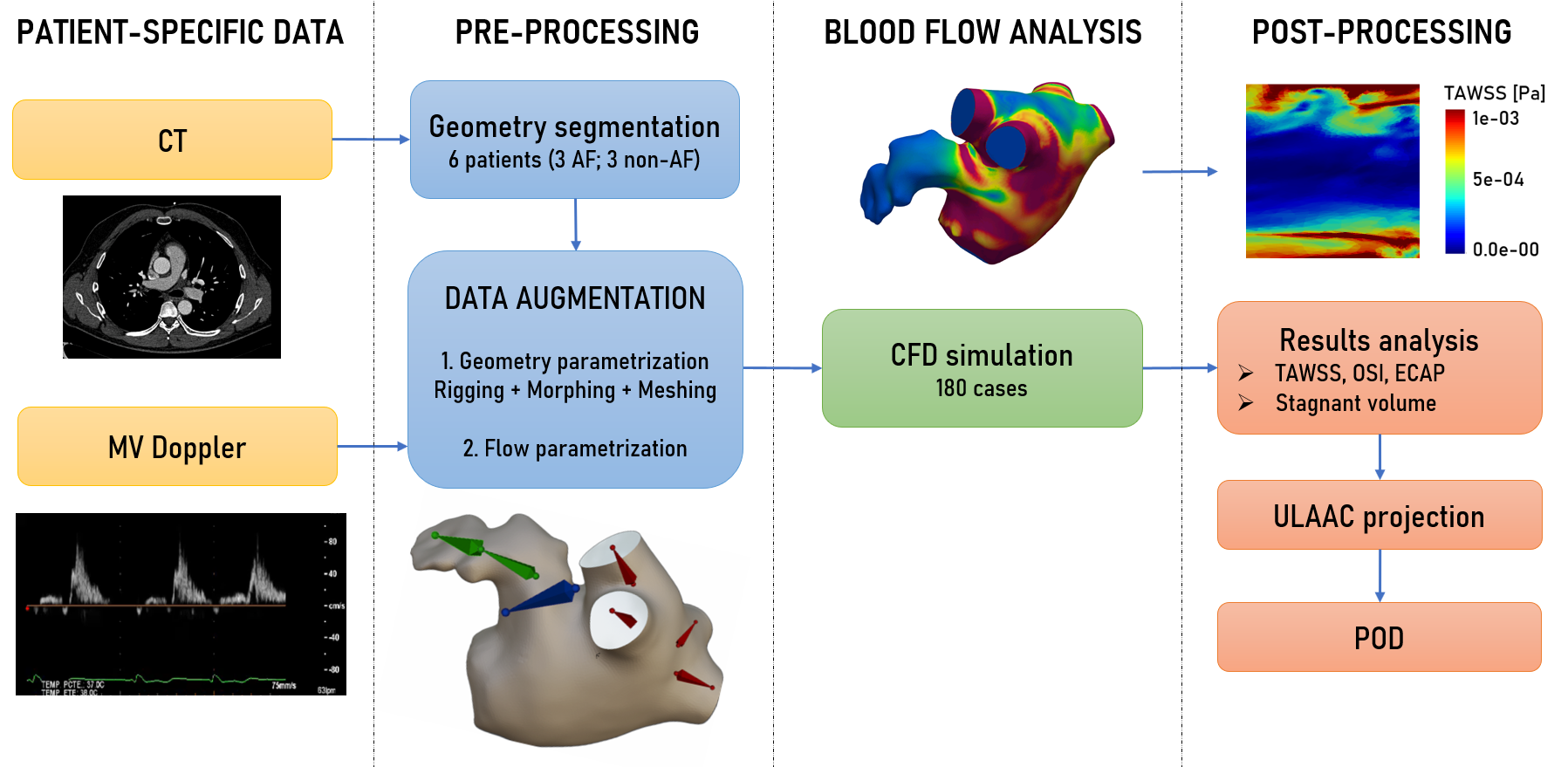}}
\caption{Workflow of the parametric blood flow analysis. \textit{Patient-specific data:} \rebuttal{computed tomography (CT)} imaging and \rebuttal{mitral valve (MV)} echo-Doppler. \textit{Pre-processing:} After geometry segmentation, a data-augmentation step was performed by rigging and morphing each of the patient-specific geometries \cite{duenas2022morphing}. \textit{Blood flow analysis:} \rebuttal{computational fluid dynamics (CFD)} simulations were performed on the data-augmented database with different volumes of \rebuttal{the left atrial appendage (LAA)}, ostium areas, and cardiac outputs to test the influence of each variable on the different geometries considered. \textit{Post-processing:} Projecting the simulation results in \rebuttal{a universal atrial appendage coordinates (ULAAC)} system and applying \rebuttal{proper orthogonal decomposition (POD)} allowed us to obtain meaningful flow features associated with LAA stasis.}
\label{fig:workflow}
\end{figure}

We imaged six human subjects by computed tomography (CT) covering the LA and LAA. Subsequently, we performed image segmentation, LAA morphing to augment the subject database, and CFD simulations on the augmented database. The simulation results were projected \rebuttal{into a universal left atrial appendage coordinate (ULAAC) system} to feed a proper orthogonal decomposition algorithm (POD). We leveraged the POD results to generate a reduced-order model to elucidate representative flow patterns and cluster patients. This workflow is summarized in Figure \ref{fig:workflow}.  

\subsection{Medical imaging}
\rebuttal{We retrospectively studied 6 patients using CT static imaging: three patients with AF  provided by the Cardiology Service of the University Hospital of Badajoz, Spain (AF1, AF2, AF3), and three cardiac patients without AF or thromboembolic history were provided by the Cardiology Service of the Puerta de Hierro Hospital, Spain (noAF1, noAF2, noAF3)}. The imaging was performed \rebuttal{according to} standard clinical protocols in each participating center\rebuttal{, and the} data collection was approved by the Ethics Committee of each hospital. \rebuttal{AF1 and AF2 suffered from permanent AF, while AF3 suffered from recurrent paroxysmal AF.}

\rebuttal{The images were acquired with a resolution of 512x512 pixels for both AF and non-AF anatomies, with two different scanner models}: three cardiac images were acquired with a Light Speed VCT General Electric Medical Systems (Milwaukee, WI, USA) by the University Hospital of Badajoz, while the other three were obtained \rebuttal{with} a SIEMENS Sensation scanner by the Puerta de Hierro Hospital. Both scanners were 64 detector devices equipped with snapshot segment mode technology to obtain cardiac images without losing clarity. A contrast dose was injected for each patient according to standard clinical protocols. Images were taken during fibrillation for patients with AF and during sinus rhythm for patients without AF.

DICOM files were reconstructed from cardiac images using standard manufacturer algorithms, which were processed using open-source software 3D-Slicer to obtain semiautomatic 3D segmentations of the endocardial surface based on signal intensity thresholding \cite{duenas2021comprehensive}. These \rebuttal{segmentations} were cleaned with MeshMixer (Autodesk), removing the pulmonary veins (PV) geometry beyond the first bifurcation to obtain the LA/LAA geometry for each patient. The movement of the atrial wall was almost negligible for \rebuttal{AF anatomies}, so only the geometry at a time instant was selected for analysis. In contrast, for \rebuttal{non-AF anatomies, the diastolic geometries} were considered. A radiologist carefully supervised and validated this procedure to obtain six CFD-ready 3D models. 

\subsection{Atrial model rigging and morphing}

The 3D segmented models were rigged and skinned to parameterize the atrial surface as previously described \cite{duenas2022morphing}, allowing us to apply scaling, rotations, and displacements to different parts of the models. To this end, we used Blender, an open-source 3D suite with complete modeling, rigging, and morphing capabilities. 
The first step was the rigging process, which provided each 3D endocardial model with an armature constituted of different bones. Each bone controlled the mesh vertices belonging to its zone of influence: scaling, rotating, etc.
\rebuttal{We applied the \textit{envelope technique} \cite{blender_manual}, which calculates an influence metric for each bone based on its distance to surrounding points using the \textit{heat method} \cite{crane2017heat}}. Finally, these influences were assigned as weights in the vertex groups. 

\begin{figure}[t]
\centerline{\includegraphics[width=0.9\linewidth]{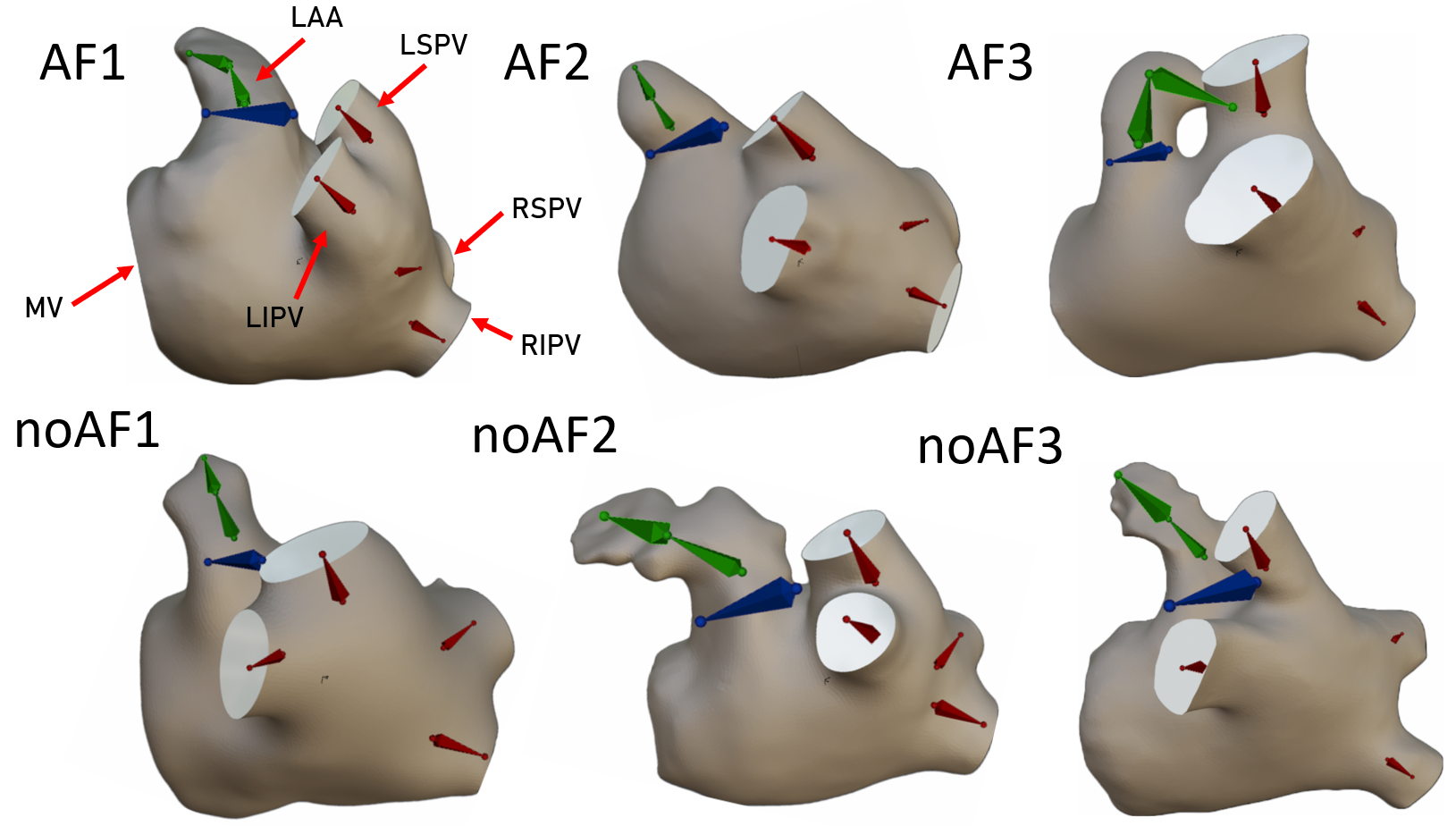}}
\caption{Endocardial model rigging. These bones made it possible to apply plausible variations to the ostium area and LAA volume while keeping the PV orientations constant throughout the 180 generated cases. LAA bones are displayed in green, ostium bones in blue, and PV bones in red.}
\label{fig:rigging}
\end{figure}

For each anatomical model, we defined the following bones: left superior PV (\rebuttal{LSPV}), left inferior PV (\rebuttal{LIPV}), right superior PV (\rebuttal{RSPV}), right inferior PV (\rebuttal{RIPV}), ostium and LAA. Each bone allowed manipulation of position, length scaling, and rotation, transferring these transformations to the mesh nodes under its influence. \rebuttal{Figure \ref{fig:rigging} shows the bones of the LAA for the six atrial segmentations. We used this approach to modify the geometrical factors that are likely to affect LAA blood stasis when the LA undergoes adverse remodeling, i.e., ostium area and LAA volume \cite{shirani2000structural, kishima2016morphologic}, while at the same time keeping PV orientations constant to eliminate biases \cite{duenas2022morphing, mill2023role}.}

\begin{figure}[t]
\centerline{\includegraphics[width=0.9\linewidth]{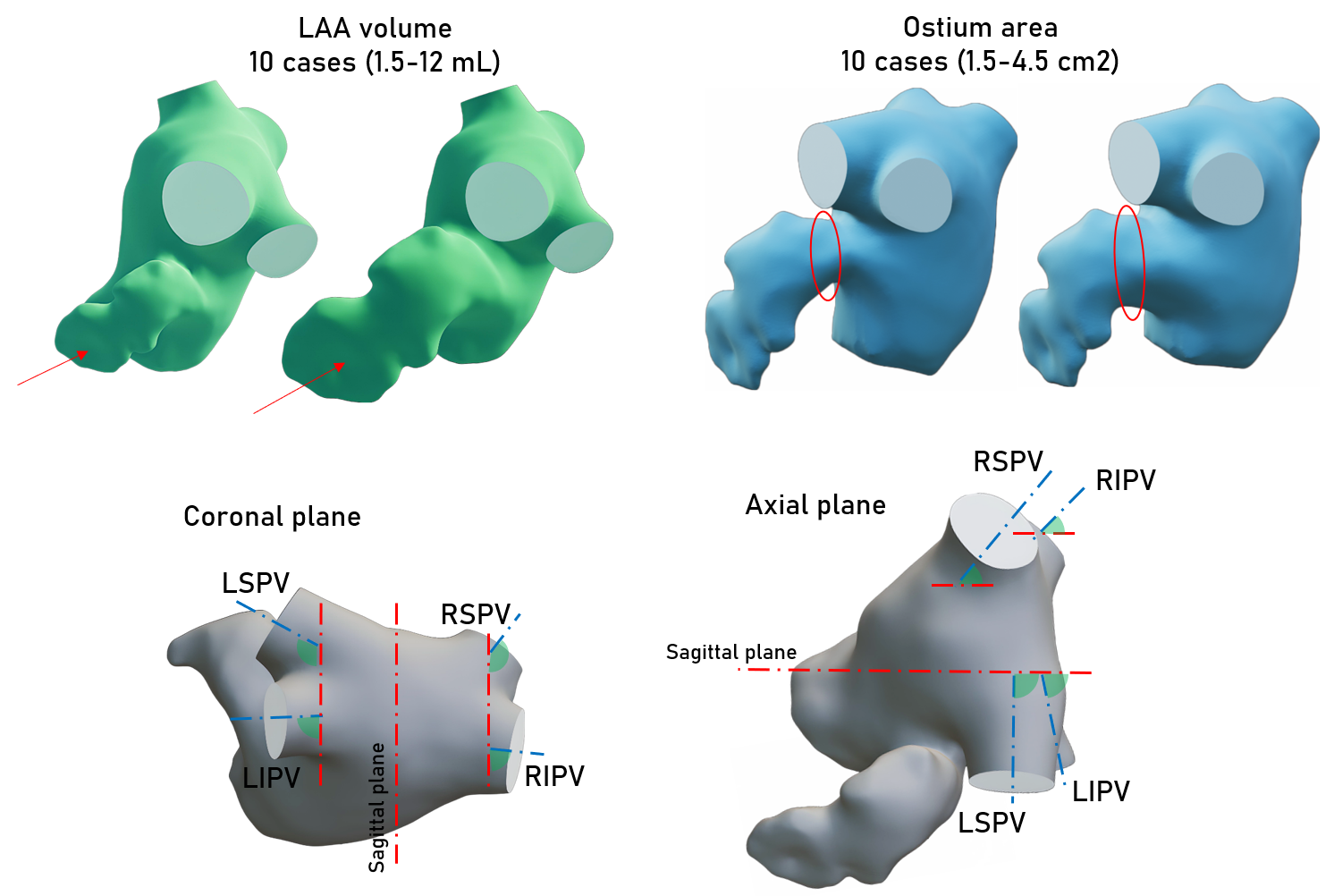}}
\caption{20 different cases were generated from each of the six geometry models; 10 cases varied the volume of the LAA (extreme cases for one of the patients presented in green), and 10 cases varied the area of the ostium (the same as before presented in blue). The projective angles were kept constant and with the same value for all generated cases: a graphical definition of the coronal and axial projections can be seen at the bottom of the figure.}
\label{fig:geometries}
\end{figure}

We generated 10 different cases for each geometric model by scaling the LAA bones to vary the LAA volume from 1.5 to 12 ml and 10 more cases by scaling the ostium bone to vary its area from 1.5 to \SI{4.5}{cm^2}. This means 20 different models for each of the six patient-specific segmentations, i.e., a total of 120 different atrial models. These values \rebuttal{are within the clinically observed range for adverse atrial remodeling \cite{ernst1995morphology}. The projective angles \cite{jue2009study, buist2016association} were kept constant across different cases to ensure a consistent orientation of the PVs and eliminate biases in the analysis. That is, the angles between each PV's coronal and axial projections and the sagittal plane were kept constant throughout the 120 different cases}. The graphical definition of each projective angle can be seen in Figure \ref{fig:geometries}.

\subsection{Atrial CFD simulations}

We used OpenFOAM open-source software to run CFD simulations. The simulation used the OpenFOAM pimplefoam application, a transient and incompressible method 
\rebuttal{that combines} the PISO and SIMPLE algorithms. An implicit Euler time integration scheme was used, while the spatial discretization employed a cell-limited Gaussian linear gradient scheme and a linear upwind divergence scheme. A preconditioned biconjugate gradient solver (PBiCGStab) was used to solve the velocity field and the passive scalars, with a convergence criterion of $10^{-5}$ for the residuals.

The fluid was considered to have the same density and viscosity as blood, i.e. $\rho = \SI{1050}{kg/m^3}$ and $\mu = \SI{0.0035}{\pascal\cdot\second}$ respectively \cite{duenas2021estimation}, and it was assumed to behave as a Newtonian fluid. 
The flow was computed by solving the continuity and Navier-Stokes equations (Equations \ref{eq:continuity} and \ref{eq:momentum}, respectively) where $\textbf{v}$ is the velocity vector and $p$ is the pressure,

\begin{equation} \label{eq:continuity}
\nabla \cdot \textbf{v} = 0
\end{equation}

\begin{equation} \label{eq:momentum}
\frac{\partial{\textbf{v}}}{\partial{t}} + \textbf{v} \cdot \nabla \textbf{v} = - \frac{ \nabla p}{\rho} + \frac{\mu}{\rho} {\nabla}^2 \textbf{v}
\end{equation}\\

A tetrahedral mesh was constructed for each case using the OpenFOAM tetMesh application. For each case, a residual and mesh independence analysis was performed to validate the flow and to verify that the velocities and pressures converged and did not depend on the mesh size. Each final mesh consisted of approximately 800k cells, varying slightly from case to case. 

\rebuttal{The atrial walls were held fixed during the simulation for all cases. Therefore, our non-AF simulations represent normal LA anatomy but not normal LA function. 
This simplification allowed us to dissect the geometric and functional effects of atrial remodeling.
To highlight this feature, we labeled the simulations performed on non-AF patient-specific segmentations as non-AF anatomy 1-3 and those performed on AF segmentations as AF anatomy 1-3.
} 
A single geometry was taken for patients with AF, \rebuttal{while the diastolic geometry was taken for non-AF anatomies. Diastolic geometry} was chosen because it is more critical than systole in terms of stasis within the LAA: At equal cardiac output, stasis while performing a rigid-wall simulation with diastole geometry is higher because the LA/LAA volume is larger at this point of the cycle.

Each cardiac cycle lasted $T=1.08$ s, and the time step was adjusted to keep the CFL number below 0.1. A time convergence analysis demonstrated that the simulation lost memory of the initial condition after four simulation cycles, so the results presented in the following section are based on the fifth simulated cycle. 

The patient-specific transmitral flow profile used in \cite{duenas2022morphing} was scaled to vary the cardiac output from 2.5 to 7.5 L/min. This profile did not have the typical A-wave of atrial contraction to replicate the AF conditions \cite{Bosi2018}. \rebuttal{This total transmitral flow rate was evenly distributed among the four PVs to establish the inflow boundary conditions \cite{duran2023pulmonary, duenas2022morphing}, thus dividing by four the total transmitral flow}. A constant pressure equal to zero was established as the boundary condition of the outlet \cite{duenas2021comprehensive}. Together with the 20 cases per segmentation obtained by rigging and morphing, the 10 additional cases obtained by varying the flow rate led to a total of 180 simulations for the six patient-specific segmentations. This procedure allowed us to study parametrically how each selected variable affects LA flow and LAA stasis for each patient-specific geometry. All variables present in parameterization and their ranges of variation can be seen in Table \ref{tab:simulations}.

\begin{table}[t]
\centering
\caption{Parameters, ranges of variation, and number of simulated cases.}
\begin{tabular}{cc|c|}
\hline
\multicolumn{1}{|c|}{\textbf{Parameter}} & \textbf{Range} & \textbf{N cases}                \\ \hline
\multicolumn{1}{|c|}{LAA volume}     & 1.5-$\SI{12}{mL}$     & 10 variations x   6 patients =   60 cases \\ \hline
\multicolumn{1}{|c|}{Ostium area}    & 1.5-$\SI{4.5}{cm^2}$   & 10 variations x 6   patients =   60 cases \\ \hline
\multicolumn{1}{|c|}{Cardiac output} & 2.5-$\SI{7.5}{L/min}$ & 10 variations x 6   patients =   60 cases \\ \hline
& & \textbf{TOTAL: 180 cases} \\ \cline{3-3} 
\end{tabular}
\label{tab:simulations}
\end{table}

\subsection{Hemodynamic indices}
\label{sec:indices}

We analyze blood age moments \cite{Sierra2017} to determine stagnant regions with a high residence time (RT).   Similar to our previous works, \cite{duenas2021comprehensive, duenas2022morphing}, we define the $k$-th age moment as
\begin{equation}\label{eq:moment_def}
  m_k = \int_{0}^{\infty}{t^k f(t)\, dt}
\end{equation}
where $f(t)$ is the age distribution of the blood and $t$ is simulation time. \rebuttal{Being the function f(t) the age distribution, m1 is the mean age (or RT of the fluid), m2 is the age variance, m3 the age skewness, m4 the age kurtosis, etc.}
The moment equations of the age distribution were integrated into the solver by including a set of scalarTransport functions in the OpenFOAM case-control dictionary. 

We computed the first moment of the distribution function ($m1$), that is, the RT of the fluid, as well as its standard deviation $ 
\sigma = \sqrt{m_2 - m_1^2}$. We then used these variables to define the normalized first moment
  \begin{equation}
    M_1 = \frac{m_1}{\sigma},
  \end{equation}
which has a bimodal distribution inside the LAA \cite{duenas2021comprehensive}. This bimodality allows for automatically selecting a patient-specific threshold to delineate stagnant volumes, that is, the value of $M_1$ corresponding to its minimum probability valley between the two modes.

We also calculated wall shear-based indicators, such as the time-averaged wall shear stress (TAWSS)\rebuttal{,  the oscillatory shear index (OSI), and the endothelial cell activation potential (ECAP)}. The formal definition of TAWSS is

\begin{equation}
TAWSS = \frac{1}{T} \int_0^T \lvert WSS \rvert dt.
\end{equation}

This hemodynamic indicator is commonly used since low TAWSS values are associated with endothelial cell damage that activates the extrinsic coagulation cascade \cite{chiu2011effects}. On the other hand, the OSI is a dimensionless indicator between 0 and 0.5 which captures the wall-shear direction oscillations and which definition is:
\begin{equation}
OSI = \frac{1}{2} \left(1 - \frac{\lvert \int_0^T  WSS dt \rvert }{\int_0^T \lvert WSS \rvert dt} \right).
\end{equation}

\rebuttal{Lastly, the ECAP is a hemodynamic indicator determined as the OSI divided by the TAWSS. High ECAP values are linked with high endothelial susceptibility and thrombogenesis risk \cite{Achille2014}. ECAP was recently applied to assess the risk of thrombosis within the LAA \cite{mill2021sensitivity, morales2021deep, zhang2023computational}.}

\subsection{Universal left atrial appendage coordinate projection}
\label{sec:ulaac}

We developed a universal left atrial appendage coordinate (ULAAC) system that mapped different LAA surfaces onto a common plane domain to facilitate data visualization and compare simulation results from different anatomical models. The mapping was inspired by the universal atrial coordinate (UAC) system developed by Roney et al. \cite{roney2019universal}, which was preceded by the universal ventricular coordinates (UVC) by Bayer et al. \cite{bayer2018universal}. 

\begin{figure}[t!]
    \centering
   \begin{subfigure}[t]{0.44\textwidth}
        \centering
        \includegraphics[width=\linewidth]{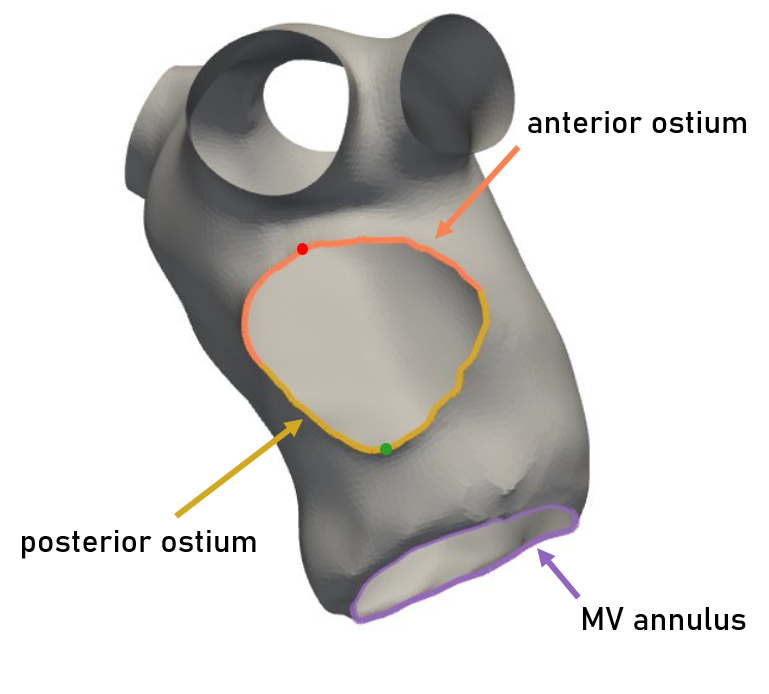}
        \caption{}
          \label{subfig:ulaac_bcs_a}
    \end{subfigure}
    \begin{subfigure}[t]{0.52\textwidth}
        \centering
        \includegraphics[width=\linewidth]{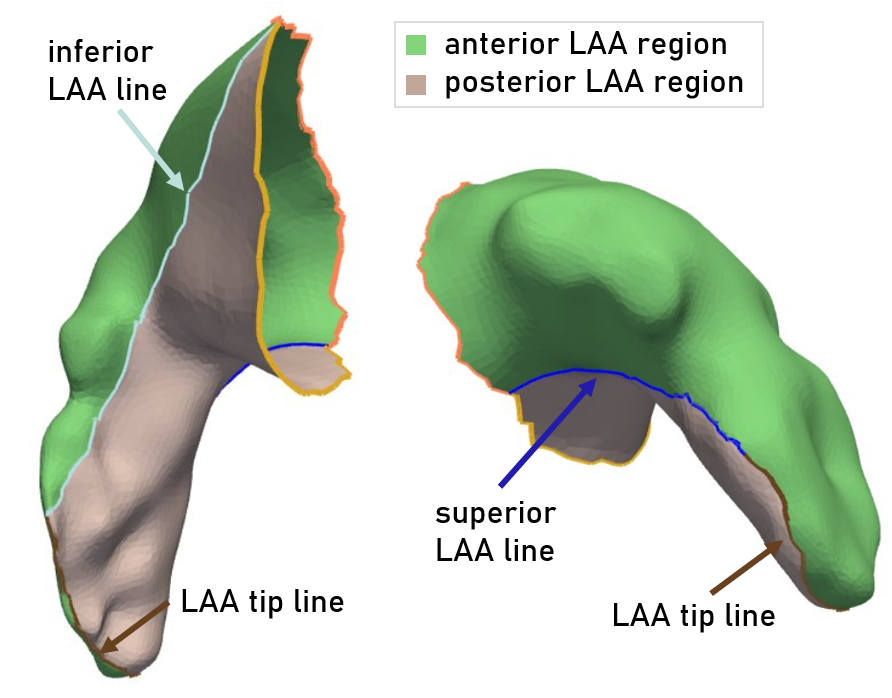}
        \caption{}
          \label{subfig:ulaac_bcs_b}
    \end{subfigure}    
\caption{Only the ostium and the LAA tip line are user-defined; the definition of the other boundary lines is automated. (a) \rebuttal{The closest and farthest ostium points of the MV annulus (marked in green and red, respectively)} were used to divide the ostium into two segments, \rebuttal{the midpoints of which divided the ostium into anterior and posterior segments}. (b) The intersections between \rebuttal{the} anterior and posterior ostium were connected to the ends of the LAA tip line to define the inferior and superior LAA lines. These lines delimited the anterior and posterior regions of the LAA.} 
\label{fig:ulaac_bc}
\end{figure}

\rebuttal{A pair of coordinates was defined on each LAA surface, effectively mapping all LAA shapes into the unit square. The superior-inferior $\delta_{LAA}(x,y)$ and the anteroposterior $\gamma_{LAA}(x,y)$ directions were chosen to ensure that the ULAAC isolines were approximately orthogonal to each other on the LAA surface. To calculate each coordinate, a Laplace equation was solved in the chosen direction, prescribing zero and unit Dirichlet boundary conditions along a pair of boundary lines corresponding to anatomical landmarks. The boundary lines chosen to solve the superior-inferior $\delta_{LAA}(x,y)$ coordinate were the superior and inferior edges of the LAA (dark and light blue lines in Figure \ref{fig:ulaac_bc}, respectively). In contrast, the ostium (orange and yellow lines) and the LAA tip (brown line) were selected to solve the anteroposterior $\gamma_{LAA}(x,y)$ coordinate.}

\rebuttal{The procedure to define the mentioned boundary lines was as follows. The ostium line and the LAA tip line were defined as the shortest geodesic path between some user-defined points. The MV annulus (purple line in Figure \ref{fig:ulaac_bc}) was identified as the largest list of connected atrial edges. The minimum geodesic distance \cite{crane2013geodesics} was then calculated between each node belonging to the ostium line and the nearest node belonging to the MV annulus. The ostium points nearest and farthest to the MV annulus (green and red points in Figure \ref{subfig:ulaac_bcs_a}) were used to divide the ostium contour into two auxiliary sections, whose midpoints were used to divide the ostium into anterior and posterior segments (orange and yellow lines in Figure \ref{subfig:ulaac_bcs_a}, respectively). Finally, the superior and inferior LAA lines were defined as the shortest geodesic paths between the anterior and posterior ostium segments and the ends of the LAA tip line (dark and light blue lines in Figure \ref{fig:ulaac_bc}, respectively).}

\rebuttal{The superior-inferior coordinate $\delta_{LAA}$ was then calculated by solving the Laplace equation $\nabla^2 \delta_{LAA} = 0$, setting the superior and inferior lines as boundary conditions ($\delta_{LAA}|_{\text{sup}}=0$, $\delta_{LAA}|_{\text{inf}}=1$). Similarly, the anteroposterior coordinate $\gamma_{LAA}$ was constructed between the previously defined ostium and the LAA tip lines. In this case, dividing the atrial mesh into its anterior and posterior regions was necessary before solving the Laplacian equation. The anterior LAA region (marked green in Figure \ref{subfig:ulaac_bcs_b}) was defined as the surface mesh contained by the superior LAA line, the tip line, the inferior LAA line, and the anterior ostium. Similarly, the posterior LAA region (marked salmon in Figure \ref{subfig:ulaac_bcs_b}) was defined as the surface mesh contained by the superior LAA line, the tip line, the inferior LAA line, and the posterior ostium. Then, the anteroposterior coordinate $\gamma_{LAA}$ was calculated by solving the Laplace equation $\nabla^2 \Psi = 0$, setting the anterior and posterior ostium as boundary conditions ($\Psi|_{\text{ant}}=0$, $\Psi|_{\text{post}}=1$).}

\begin{figure}[H]
\centerline{\includegraphics[width=0.9\linewidth]{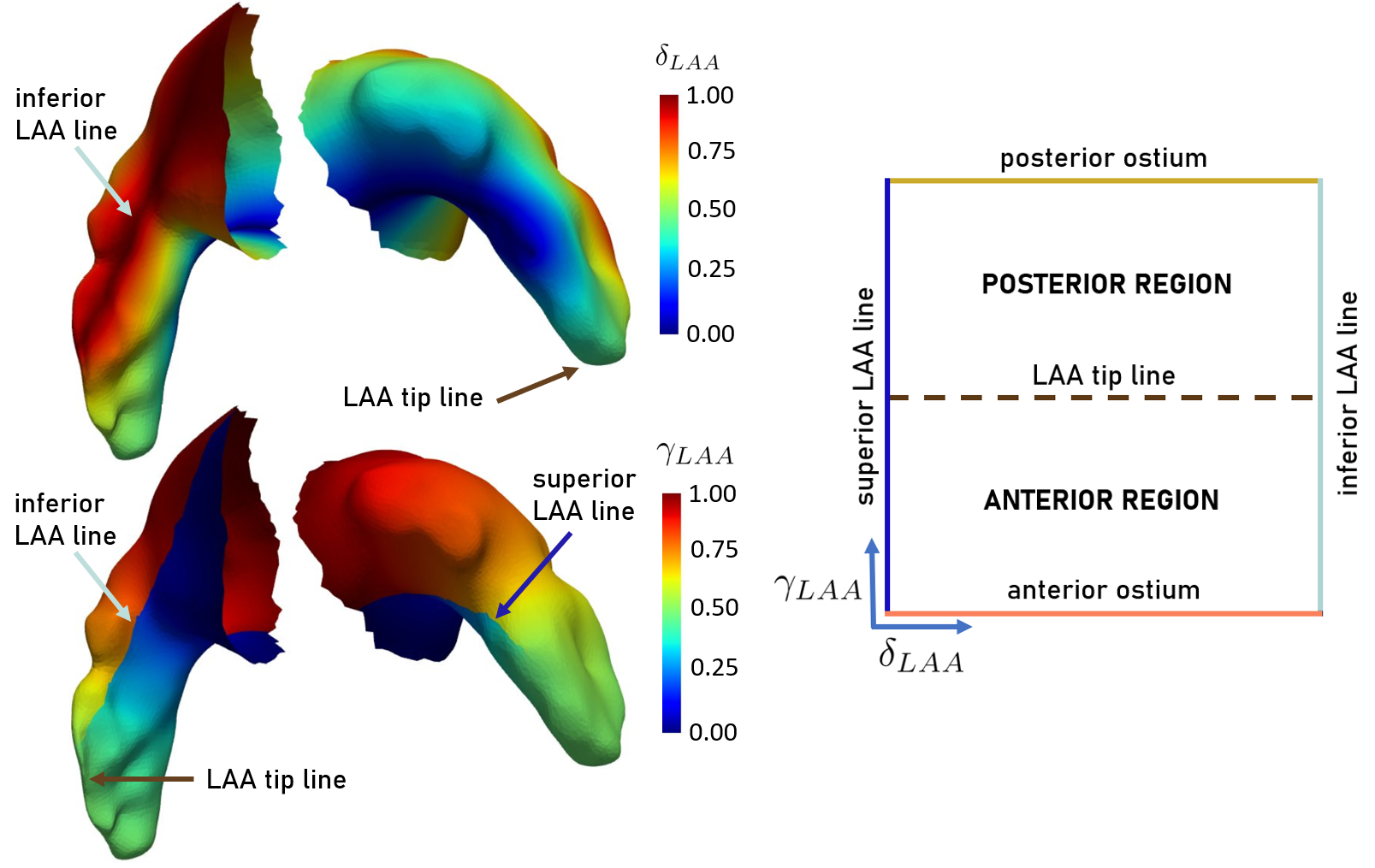}}
\caption{The Laplace solution for the LAA superior-inferior coordinate $\delta_{LAA}$ ranges from 0 on the superior LAA line to 1 on the inferior LAA line. The LAA anteroposterior coordinate $\gamma_{LAA}$ ranges between 0 and 0.5 in the LAA anterior region (from the anterior ostium to the tip) and between 0.5 and 1 in the LAA posterior region (from the tip to the posterior ostium), after transforming the Laplacian solution \rebuttal{$\Psi$}.}
\label{fig:laplace_solve}
\end{figure}

\rebuttal{Finally, to ensure a bijective correspondence between each pair of coordinates $(\delta_{LAA}, \gamma_{LAA})$ and each point on the LAA surface, the anteroposterior solution $\Psi$ obtained by solving the Laplace equation was rescaled and unwrapped so that it ranged between 0 and 0.5 in the anterior LAA region ($\gamma_{LAA} = 0.5 \cdot \Psi$) and between 0.5 and 1 in the posterior LAA region ($\gamma_{LAA} = 1 - 0.5 \cdot \Psi$). The Laplace problems were solved using Python LaPy \cite{reuter2006laplace, wachinger2015brainprint}. Figure \ref{fig:laplace_solve} shows an example of the $(\delta_{LAA}, \gamma_{LAA})$ fields represented in a patient-specific anatomical model.}

\subsection{Proper orthogonal decomposition}

We used the snapshot method \cite{krysl2001dimensional} to compute the proper orthogonal decomposition (also known as principal component analysis) of hemodynamic indices in the reference system $(\delta_{LAA}, \gamma_{LAA})$, using 100 grid points in each direction. 
Calculations were carried out with the LAPACK routine gesdd, which computes the singular value decomposition of the snapshot matrix $A_{M,N}$ where $M=10^4$ was the number of spatial grid points and $N=180$ the number of simulations.  

POD allows for expressing each patient's hemodynamic variables, e.g., \rebuttal{TAWSS, OSI or ECAP}, as 
\begin{equation}
    F(\delta,\gamma) = \sum_{n=1}^N a_n \phi_n (\delta,\gamma)
    \label{eq:POD}
\end{equation}
where $n$ is mode rank, the coefficient $a_n$ indicates each mode's contribution to the dataset variance, and $\phi_n(\delta,\gamma)$ encodes the mode's spatial structure. One of the main advantages of POD is that $a_n$ decays strongly with $n$, making this decomposition suitable for dimensionality reduction. The first four modes accounted $> 90\%$ of the variance in our database, so we considered reduced-order models with four modes.

In this work, we hypothesized that it is possible to differentiate between \rebuttal{anatomies with and without AF according to each anatomy projection on the POD modes}. 
\rebuttal{We represented each particular dataset as
$
    F^j(\delta,\gamma) = \sum_{n=1}^N a^j_n \phi_n (\delta,\gamma)
$
where $j=1,\dots, 180$ labels the specific dataset, $n$ denotes POD mode rank, and the functions $\phi_n (\delta,\gamma)$ are the POD eigenmodes. Thus, $a^j_n $ measures the projection of each hemodynamic map from the dataset $j$ into the eigenmode $n$ of the POD of the full cohort.}
Subsequently, all cases were projected on these modes, and the coefficient distributions $a_n$ were compared.

\section{Results}\label{sec:results}

\subsection{Stagnant LAA blood volumes in AF and non-AF \rebuttal{anatomies}}

\begin{figure}[t!]
    \centering
   \begin{subfigure}[t]{0.32\textwidth}
        \centering
        \includegraphics[width=\linewidth]{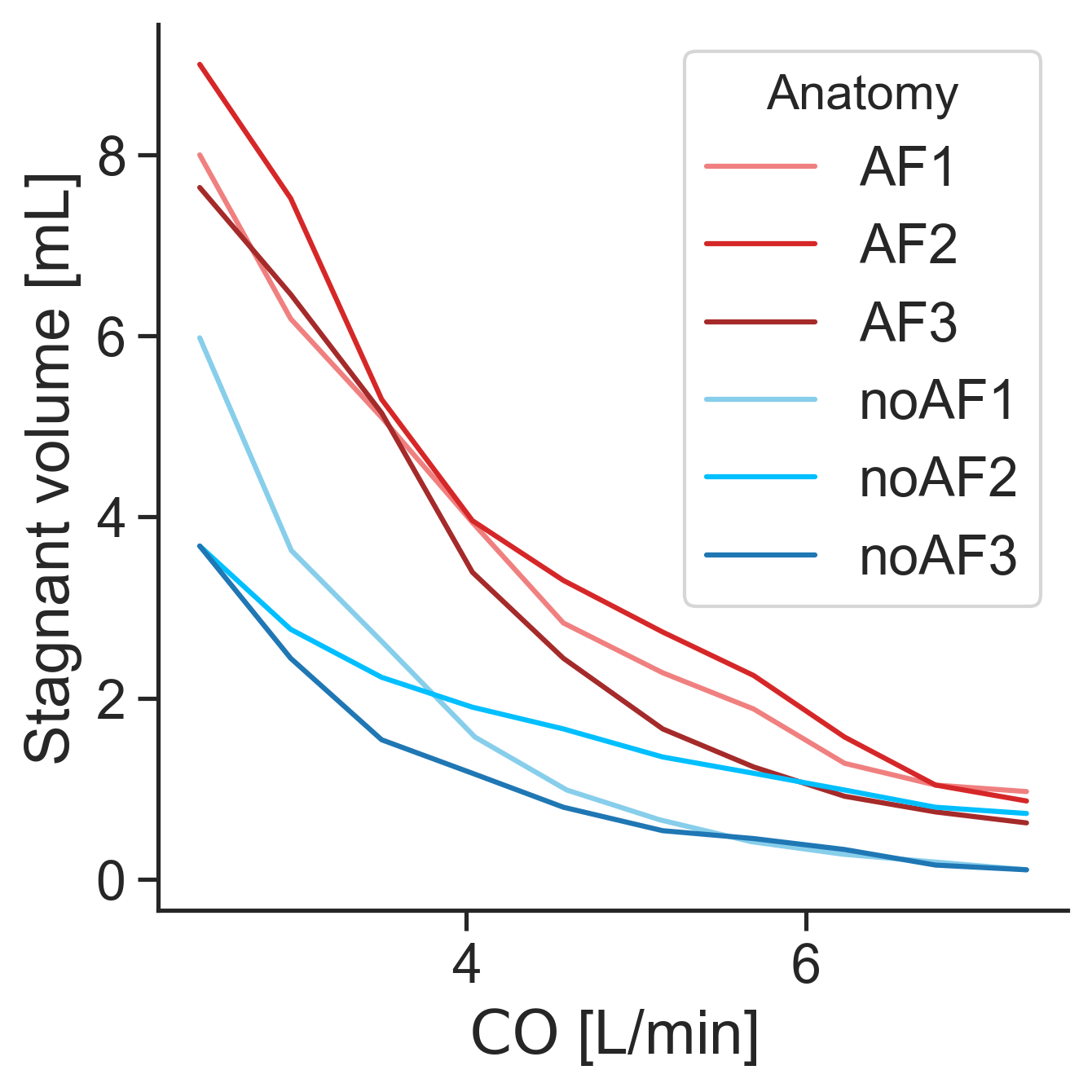}
        \caption{}
          \label{subfig:DVev_cardiac_output}
    \end{subfigure}
    \begin{subfigure}[t]{0.32\textwidth}
        \centering
        \includegraphics[width=\linewidth]{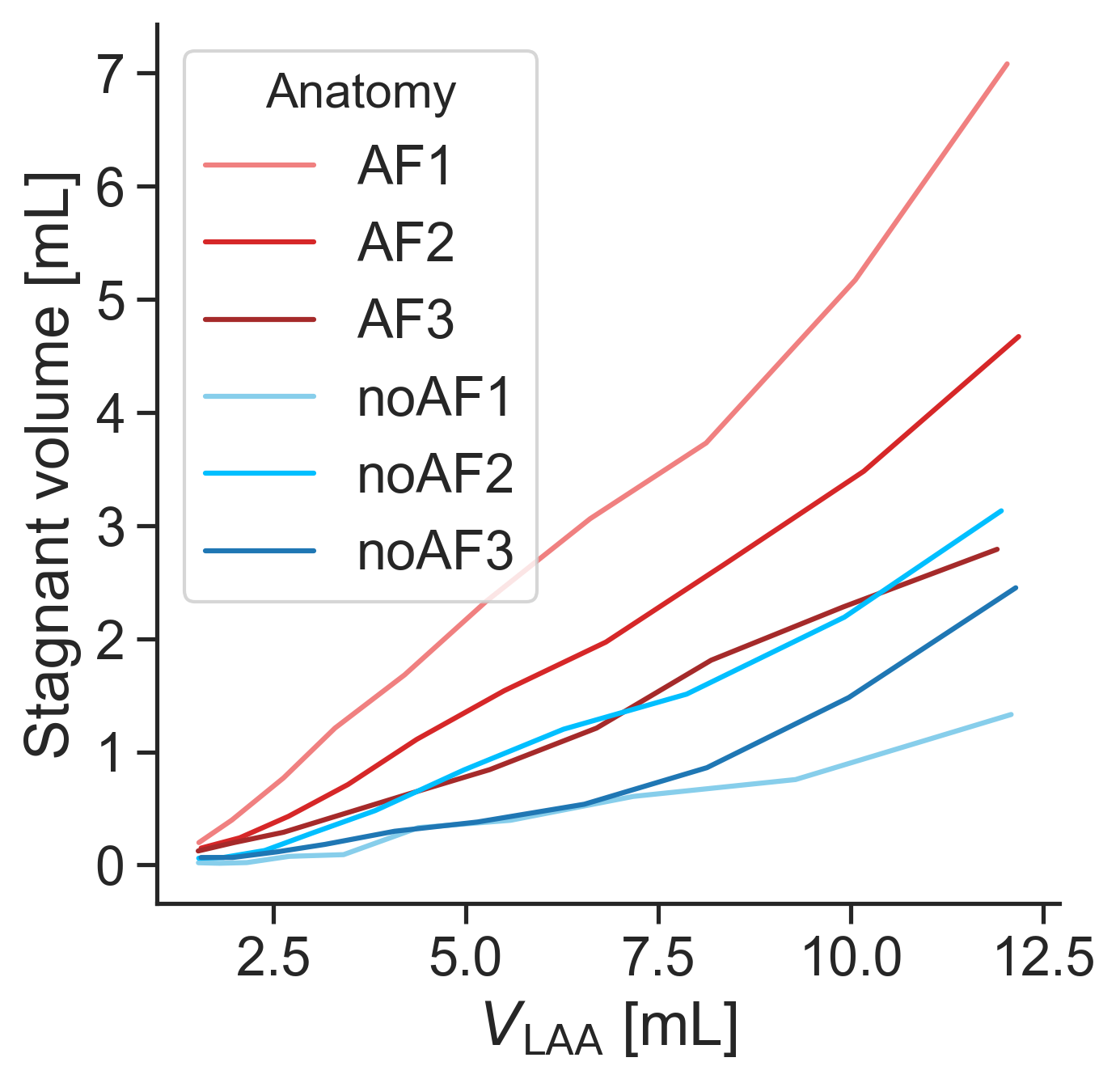}
        \caption{}
          \label{subfig:DVev_laa_volume}
    \end{subfigure}    
    \begin{subfigure}[t]{0.32\textwidth}
        \centering
        \includegraphics[width=\linewidth]{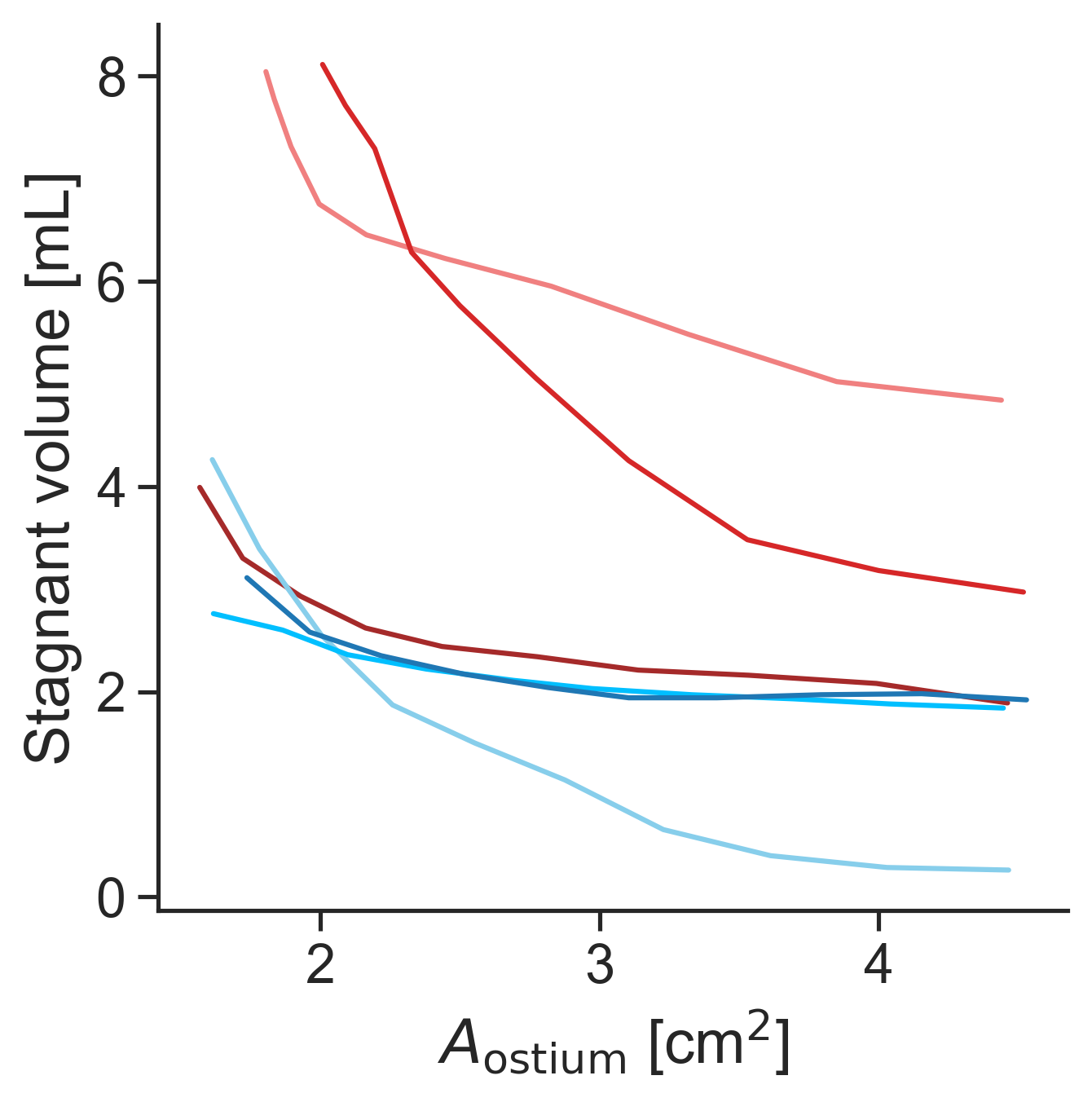}
        \caption{}
          \label{subfig:DVev_ostium_area}
    \end{subfigure}
\\
   \begin{subfigure}[t]{0.32\textwidth}
        \centering
        \includegraphics[width=\linewidth]{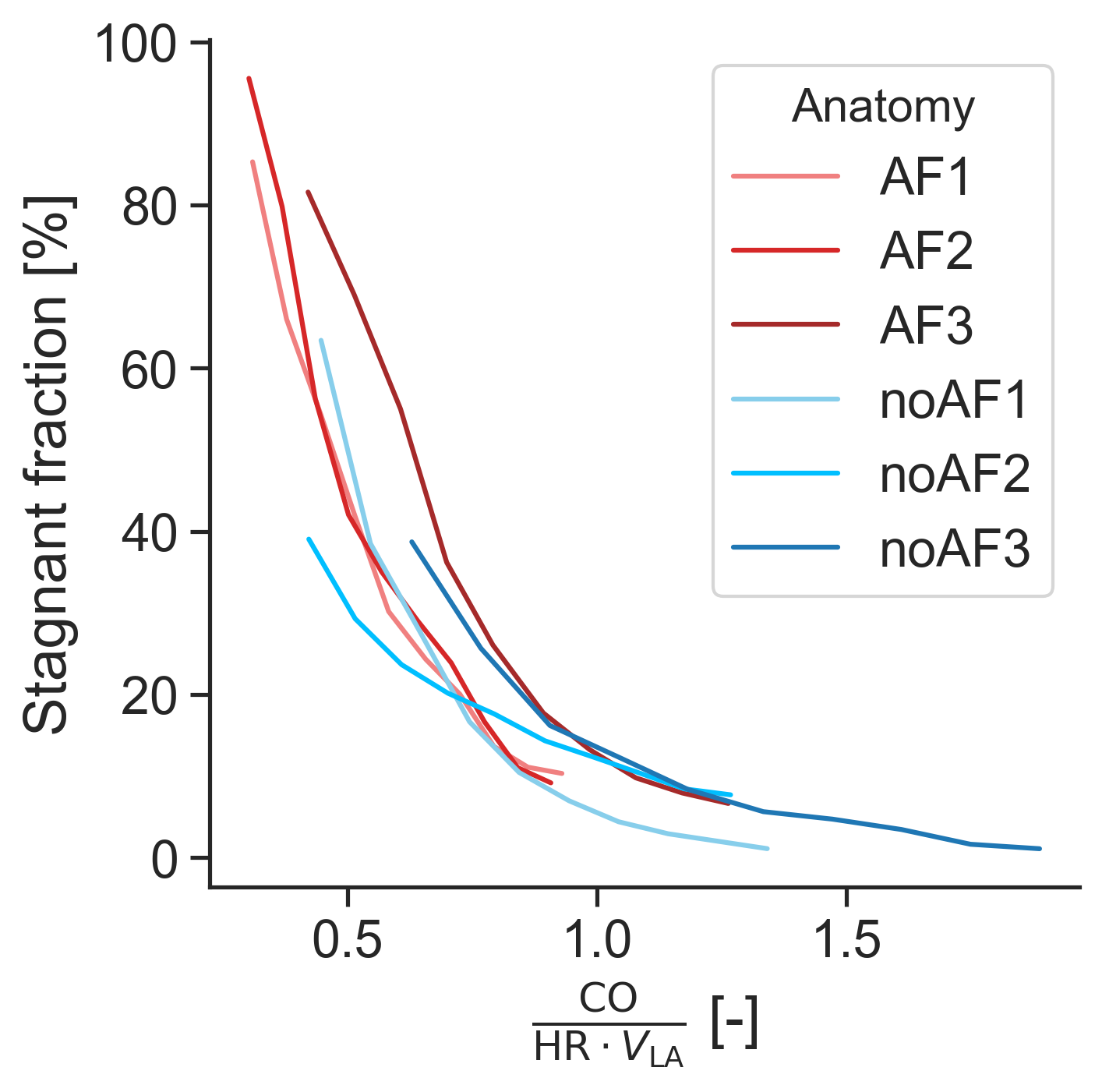}
        \caption{}
          \label{subfig:DVev_cardiac_output_adim}
    \end{subfigure}
    \begin{subfigure}[t]{0.32\textwidth}
        \centering
        \includegraphics[width=\linewidth]{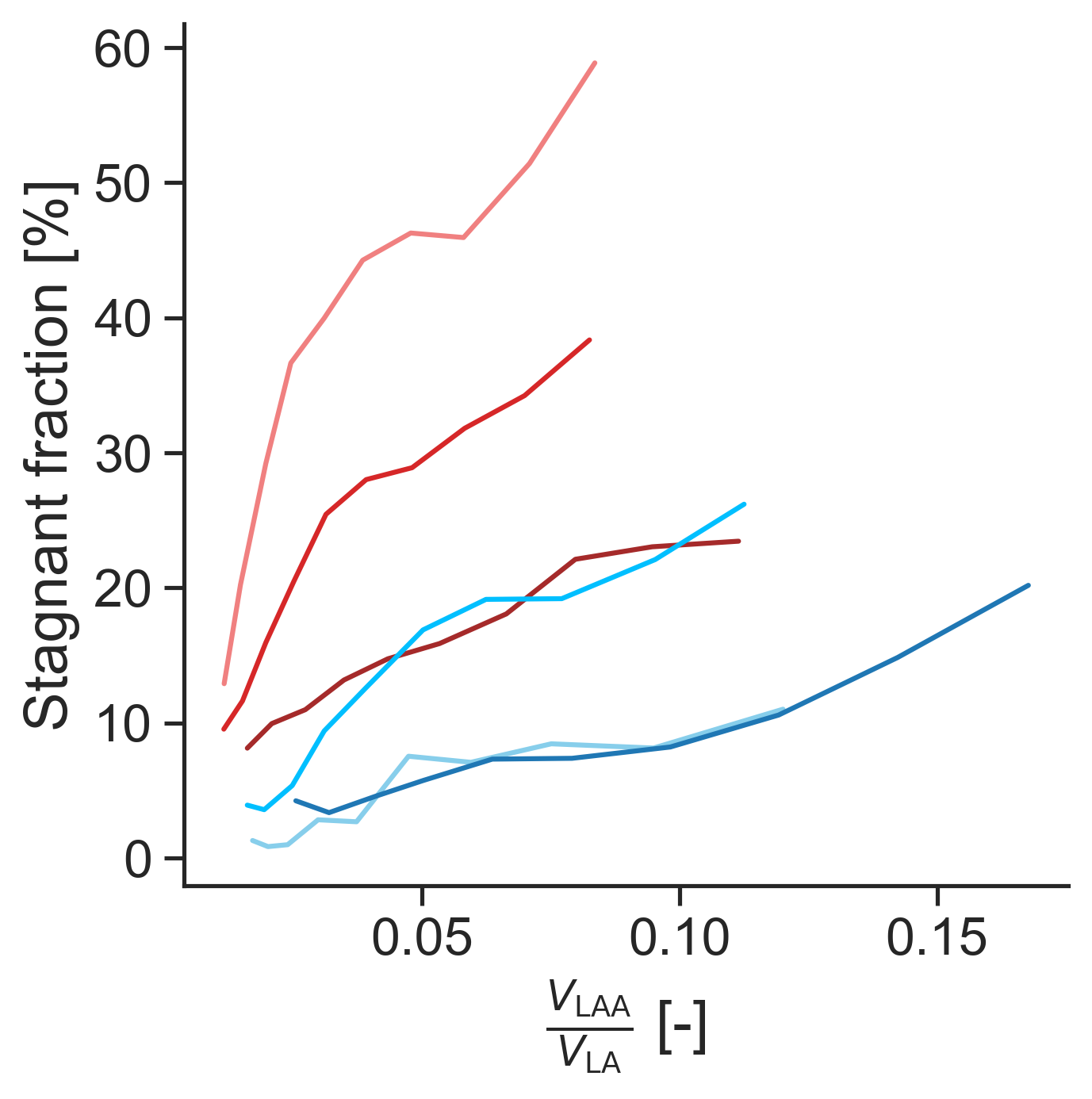}
        \caption{}
          \label{subfig:DVev_laa_volume_adim}
    \end{subfigure}    
    \begin{subfigure}[t]{0.32\textwidth}
        \centering
        \includegraphics[width=\linewidth]{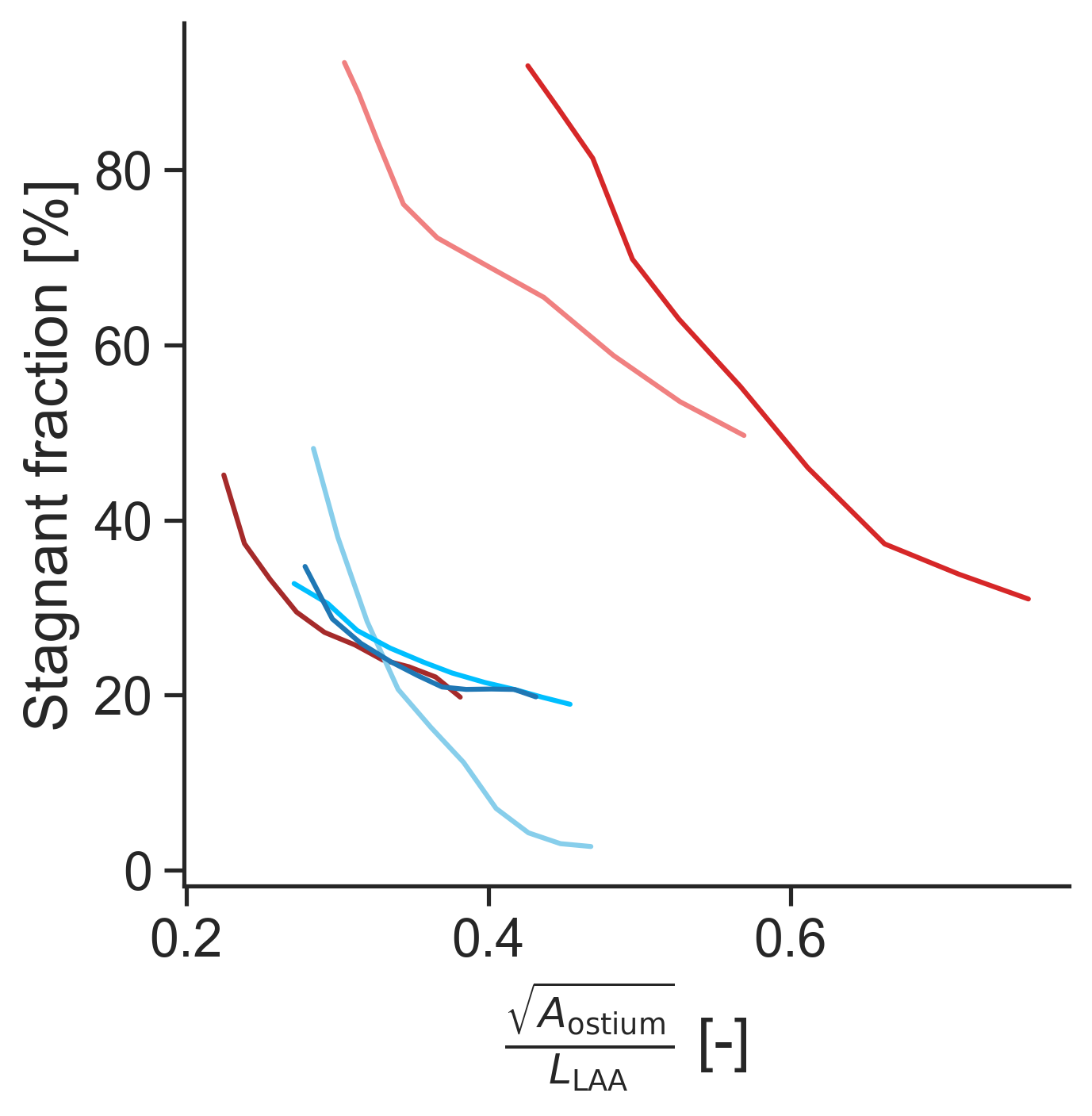}
        \caption{}
          \label{subfig:DVev_ostium_area_adim}
    \end{subfigure}
 
\caption{Evolution of the stagnant volume \rebuttal{(a, b, c) and the stagnant fraction (d, e, f)} for the 180 simulated cases. Each subfigure shows \rebuttal{the influence of varying,} respectively, (a) the cardiac output, (b) the LAA volume, (c) the ostium area\rebuttal{, (d) the non-dimensional cardiac output, (e) the non-dimensional LAA volume, and (f) the non-dimensional ostium area}. 
\label{fig:dead_volume_evolution}}
\end{figure}

We computed the stagnant volume of the 180 simulated LAAs after 5 cardiac cycles as described above. \rebuttal{Figure \ref{fig:dead_volume_evolution} displays these variables vs. cardiac output (CO, panel a), LAA volume ($V_{LAA}$, panel b), and ostium area ($A_{ostium}$, panel c). We also plotted non-dimensional versions of these quantities in the bottom row of Figure 6. Stagnant volume was normalized with LAA volume to produce stagnant volume fraction; cardiac output was normalized with heart rate and LA volume (panel d), LAA volume was normalized with LA volume (panel e), and ostium area with LAA length (panel f).}
The stagnant volume of LAA decreased sharply with cardiac output, and \rebuttal{this trend became almost the same for all patients when using non-dimensional variables}. 
The stagnant volume increased with the volume of LAA for all patients, \rebuttal{both in absolute and non-dimensional variables. This growth was pronounced for simulations on AF anatomies when compared with non-AF anatomies.} 
Finally, the stagnant volume of LAA decreased with the ostium area, \rebuttal{and this decrease was more apparent when scaling this area with LAA length to represent the effective aspect ratio of the LAA. }

\rebuttal{
While we provide a physical interpretation of these trends in the Discussion section of the manuscript, it is worth noting that the effect of cardiac output on stagnant volume seems to be more dramatic and less affected by other factors than the effect of LAA volume or ostium area.
This multifactorial sensitivity highlights} the inherent limitations of global hemodynamic metrics \rebuttal{in classifying} patient phenotypes. The following sections illustrate how reduced-order models that capture the spatial features of hemodynamic metrics could contribute to discriminating patients.

\subsection{Wall shear stress maps in the universal LAA coordinate system}

\begin{figure}[t]
\centerline{\includegraphics[width=0.99\linewidth]{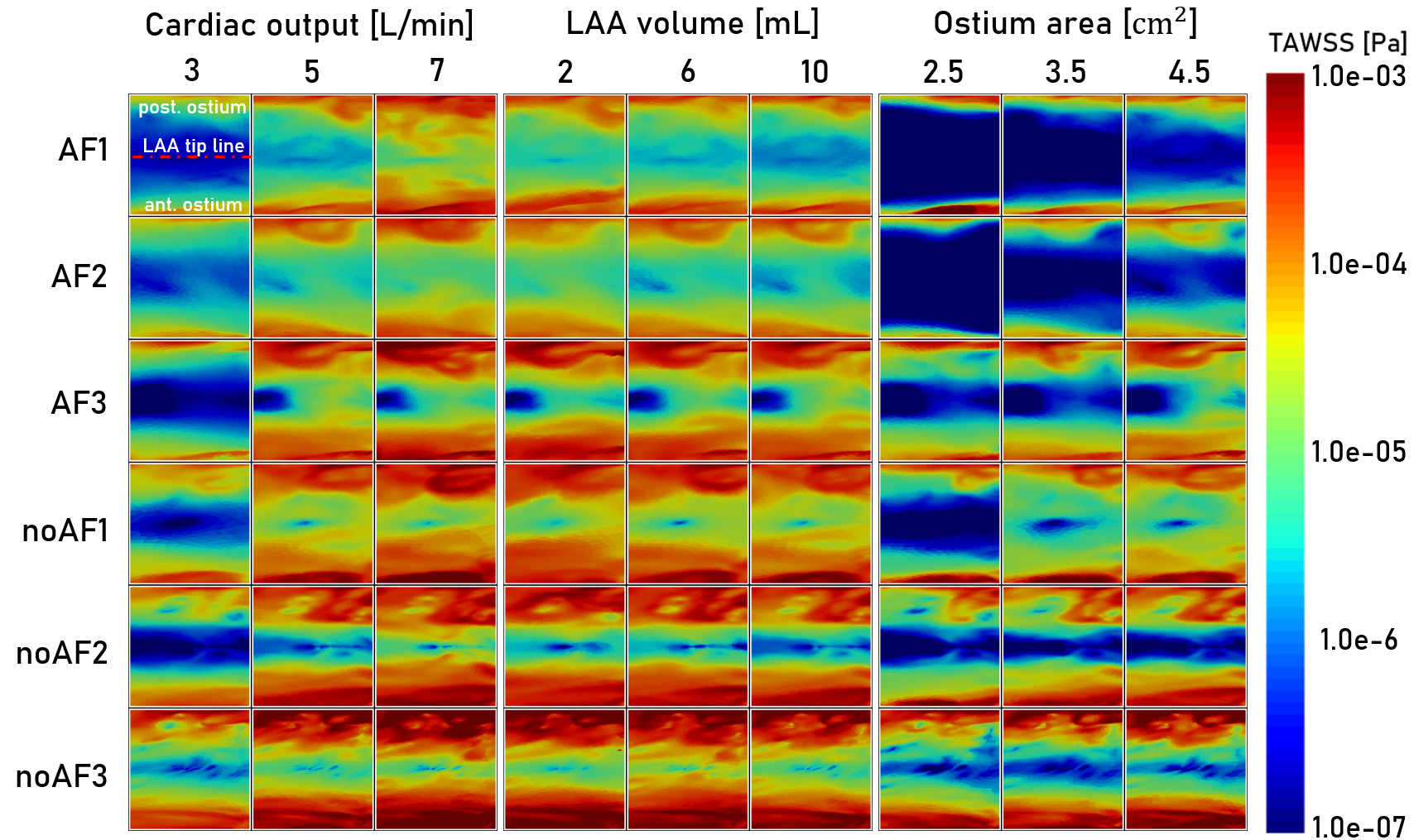}}
\caption{The TAWSS contours resulting from the simulations were projected to the unit square by employing the ULAAC coordinates, facilitating 2D visualization and data comparison between the different cases. The six rows represent the different \rebuttal{anatomies} considered. At the same time, the columns indicate the variation of TAWSS for each case by varying cardiac output, LAA volume, and ostium area, respectively.  
}
\label{fig:tawss_ulaac}
\end{figure}

\begin{figure}[t]
\centerline{\includegraphics[width=0.99\linewidth]{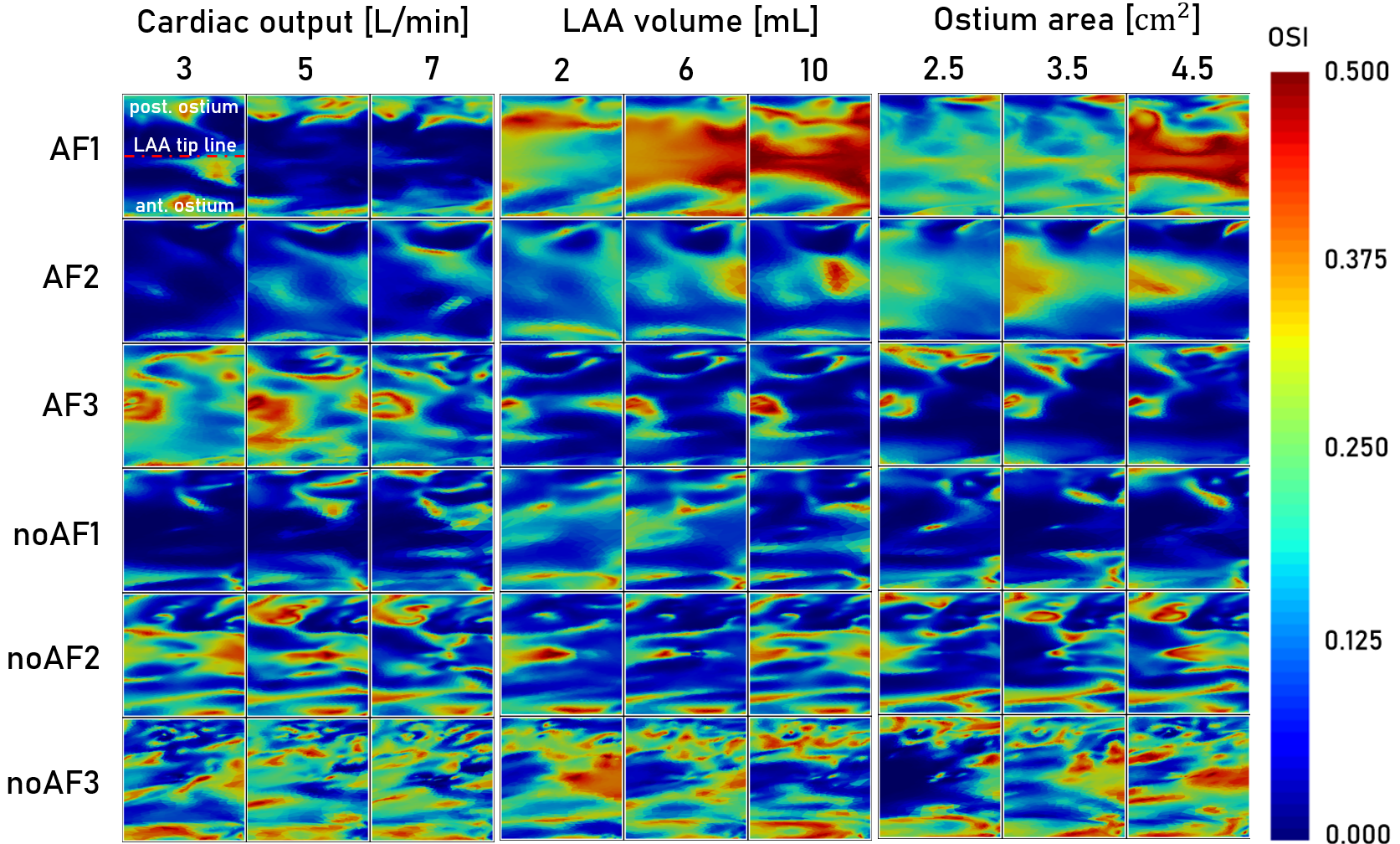}}
\caption{The OSI contours resulting from the simulations were projected to the unit square by employing the ULAAC coordinates, facilitating 2D visualization and data comparison between the different cases. The six rows represent the different \rebuttal{anatomies} considered. At the same time, the columns indicate the variation of OSI for each case by varying cardiac output, LAA volume, and ostium area, respectively.  
}
\label{fig:osi_ulaac}
\end{figure}

\begin{figure}[t]
\centerline{\includegraphics[width=0.99\linewidth]{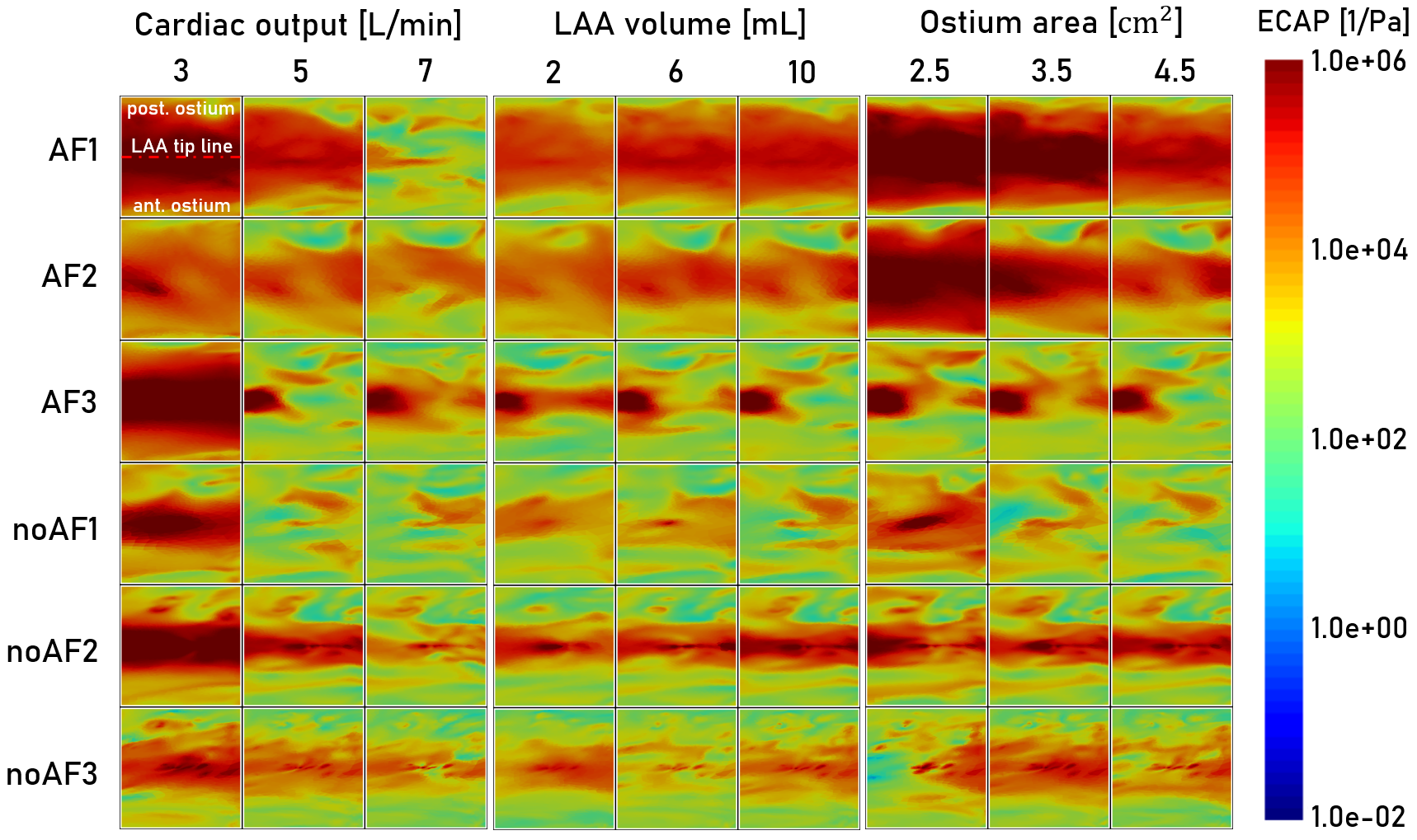}}
\caption{\rebuttal{The ECAP contours resulting from the simulations were projected to the unit square by employing the ULAAC coordinates, facilitating 2D visualization and data comparison between the different cases. The six rows represent the different anatomies considered. At the same time, the columns indicate the variation of ECAP for each case by varying cardiac output, LAA volume, and ostium area, respectively.  
}}
\label{fig:ecap_ulaac}
\end{figure}

The spatial distributions of \rebuttal{TAWSS, OSI, and ECAP} were mapped in the ULAAC system to facilitate their visualization and standardize the comparison of different simulations.
Figure \ref{fig:tawss_ulaac} shows the projected TAWSS maps, each panel representing a unique combination of baseline patient-specific segmentation (rows) and cardiac output, LAA volume, or ostium area (columns).  
These data indicate that \rebuttal{anatomies of patients with AF tend to have lower TAWSS values than patients without AF. More importantly, each patient has a characteristic TAWSS pattern that is altered by changes in cardiac output, LAA chamber volume, or ostium area altered each patient's TAWSS map.} 
 
\rebuttal{A similar analysis was performed on the OSI and ECAP maps (Figures \ref{fig:osi_ulaac} and \ref{fig:ecap_ulaac}). OSI patterns revealed that this hemodynamic metric had a more intricate spatial organization that was also more sensitive to varying simulation parameters (Figure \ref{fig:osi_ulaac}), while the ECAP maps reveal high values near the LAA tip and a high variability between patients (Figure \ref{fig:ecap_ulaac}).}
\rebuttal{In the next section, we present a systematic analysis pipeline to build reduced-order models exploiting the ULAAC system, simplifying the interpretation of these ULAAC maps from a given simulation.}

\begin{figure}[p] 
    \centering  
    \begin{subfigure}[t]{0.65\textwidth}
        \centering
        \includegraphics[width=\linewidth]{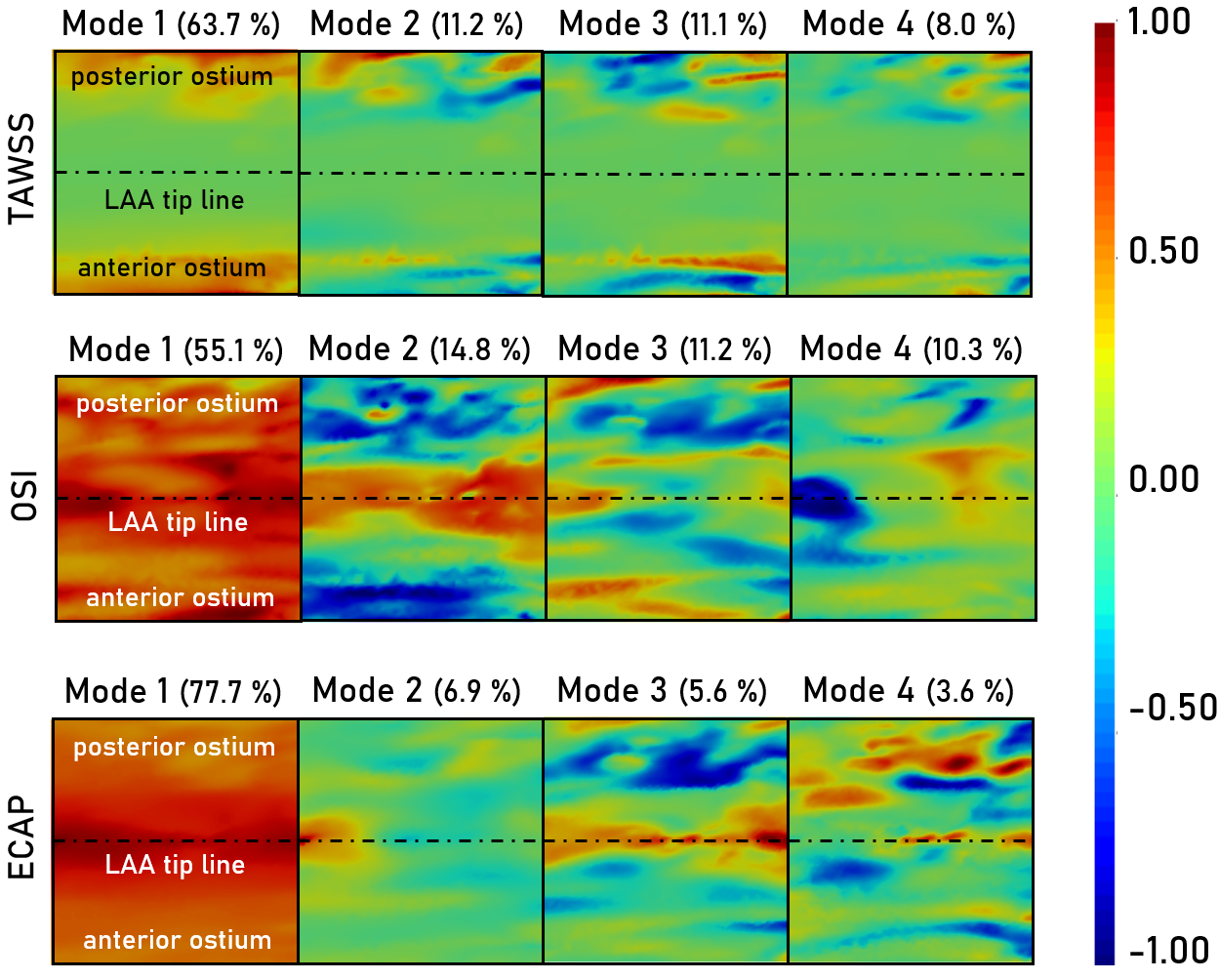}
        \caption{}
          \label{subfig:POD_modes}
    \end{subfigure}
    \begin{subfigure}[t]{0.94\textwidth}
        \centering
        \includegraphics[width=\linewidth]{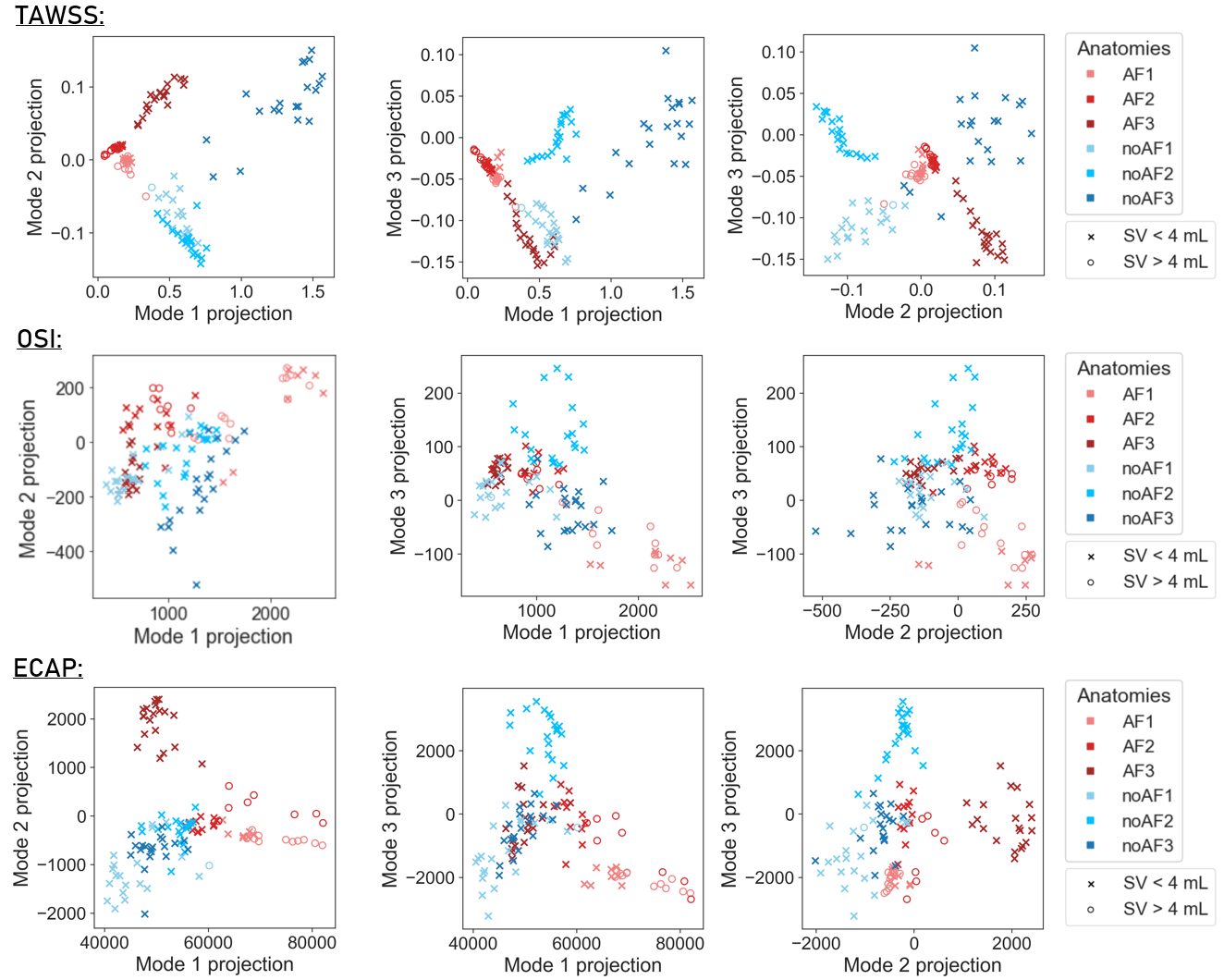}
        \caption{}
          \label{subfig:POD_projections}
    \end{subfigure}
\caption{POD was applied to the projected ULAAC results for the 180 cases to detect the underlying patterns. (a) Normalized four high-ranked modes after applying POD for \rebuttal{TAWSS, OSI, and ECAP} fields and its percentage of explained variance. (b) POD projections on the first three modes of \rebuttal{each hemodynamic indicator} for the 180 simulated cases. Different markers have been used for each case depending on whether the stagnant volume (SV) was above or below 4 mL.  \label{fig:mv_flow_bcs}}
\end{figure}

\subsection{Reduced-order model of LAA wall shear stress patterns}

Taking advantage of the structured way to represent the wall data provided by the ULAAC system, we performed a POD of \rebuttal{TAWSS, OSI, and ECAP data using the 180 simulation snapshots generated by rigging and morphing our six patient-specific LA segmentations.
The four highest-ranked POD modes accounted for more than $90\%$ of the variance of each hemodynamic parameter. Their spatial structure in the $(\delta, \gamma)$ plane ($\phi_n$ from Equation \ref{eq:POD}) was represented in Figure \ref{subfig:POD_modes}, together with each mode's percentage of explained variance ($a_n^2$ from Equation \ref{eq:POD}). The first mode represents the population's mean, whereas modes 2 to 4 encode departures from the mean of decreasing significance. }

The first TAWSS mode has near-zero magnitude along a wide band centered at the LAA tip line $\gamma = 0.5$, progressively increasing to reach a constant value as we move toward the ostium ($\gamma =0$ or $1$). Modes 2-4 \rebuttal{represent} spatial fluctuations of progressively increasing complexity near the ostium. The first OSI mode is roughly uniform, although it shows some fluctuations that could be due to our relatively low sample size of $N=6$. The second mode has a banded structure with a peak along the LAA tip line and valleys along the ostium lines. The third and fourth OSI modes encode relatively broken patterns of varying lengths and widths that are more difficult to interpret. \rebuttal{The first ECAP mode presents high values along a band centered at the LAA tip ($\gamma = 0.5$), progressively decreasing as we move towards the ostium. The second mode presents a peak in the superior LAA tip, while modes 3-4 encode spatial fluctuating patterns of varying lengths and widths.} 

Finally, we plotted $a_1^j$ vs. $a_2^j$, $a_1^j$ vs. $a_3^j$, and $a_2^j$ vs $a_3^j$ for $j=1,\dots, 180$ (Figure \ref{subfig:POD_projections}, two lower rows). These scatter plots suggest that the AF and non-AF \rebuttal{anatomies} can be segregated by jointly considering the value of two coefficients, i.e., a two-dimensional reduced order model based on POD.  
Interestingly, the point clouds \rebuttal{of the} coefficients corresponding to each patient cluster together, supporting the idea that each patient has a distinct \rebuttal{hemodynamic signature of TAWSS and ECAP} in his LAA that was not too sensitive to variations in cardiac output or geometric remodeling. 
However, the point clouds obtained from the OSI POD overlap significantly, indicating that a higher-dimensional,  reduced-order model would be required to separate AF \rebuttal{anatomies} from non-AF ones using this hemodynamic variable. This difference is consistent with the more intricate spatial structure of the OSI maps and their higher patient-to-patient variability  (Figure  \ref{fig:osi_ulaac}). 

The stagnant volume variable was also contrasted with the previously calculated POD projections, so \rebuttal{Figure \ref{subfig:POD_projections} shows with different markers the cases with a stagnant volume of less than the average stagnant volume (4 mL) versus those with a stagnant volume higher than this value.} 
In the case of TAWSS, the cases with higher stagnant volumes tend to have lower absolute values of $a_1^j$, $a_2^j$, and $a_3^j$ than the cases with lower stagnant volumes, allowing us to see how this variable is also related to the features extracted from the POD. \rebuttal{Similarly, the POD projections of the ECAP show that the cases with higher stagnant volumes are linked to high values of $a_1^j$, but not with $a_2^j$ or $a_3^j$.} On the other hand, the information that can be extracted by contrasting the stagnant volume variable with the POD projections of the OSI does not present an easily interpretable pattern. 

\section{Discussion}\label{sec:discussion}

Computational fluid dynamics (CFD) analyses of left atrial (LA) blood flow in patient-specific anatomical models are gaining recognition as a tool to investigate the hemodynamic substrate for ischemic stroke in patients with atrial fibrillation (AF) \rebuttal{\cite{bucelli2022mathematical, Otani2016, duenas2021comprehensive, Koizumi2015, corti2022impact, garcia2021demonstration, duran2023pulmonary, Mill2020, gonzalo2022non, feng2019analysis}}. 
For each prescribed combination of inflow/outflow boundary conditions and motion of the atrial wall, CFD analysis produces a faithful 4D (3D, time-resolved) representation of flow through the atrium, which allows computing surrogate metrics of thrombosis risk, for example, blood residence time, or modeling the coagulation cascade. Special attention is devoted to metrics of left atrial appendage (LAA) stasis because the LAA is the most common site of atrial thrombosis \cite{DiBiase2012, yamamoto2014complex}. However, these metrics are sensitive to parameters such as LA wall kinetics \rebuttal{\cite{duenas2021comprehensive, garcia2021demonstration}}, PV flow split \rebuttal{\cite{Lantz2018a, duran2023pulmonary}}, or cardiac output \cite{duenas2022morphing}, which can vary significantly throughout the day for a given patient. They are also sensitive to geometrical factors such as LAA volume, ostium area, or LAA morphology, which can vary over months due to adverse remodeling in patients with AF \rebuttal{\cite{boyle2021fibrosis, levy2005remodelling}}. \rebuttal{Recent studies also suggest that considering hematocrit-dependent blood viscosity and MV annulus anatomy can improve the patient-specificity of simulations \cite{gonzalo2022non, zhang2023computational}.} Furthermore, the diversity of LAA morphologies makes it challenging to interpret and compare data from CFD analyses performed on different patient-specific anatomies. Despite notable efforts, these uncertainties have made it difficult to derive a generalizable understanding of the determinants of LAA stasis \cite{morales2021deep, Chahine2023Machine}. 

\rebuttal{Our working hypothesis is that data-driven reduced-order models of hemodynamic metrics can identify distinct flow patterns associated with LAA stasis. To evaluate our hypothesis and build a methodological framework to derive such models, we} 
introduce two methodological advances. First, to model the potential effects of AF-associated atrial remodeling, we obtain patient-specific atrial anatomical models and rig and morph these models to vary the LAA volume and the ostium area \cite{duenas2022morphing}. As a result, we digitally expanded our patient database to run 180 CFD simulations on six \rebuttal{anatomies}, gaining knowledge of how atrial remodeling exacerbates each patient's hemodynamic substrate for LAA thrombosis.
\rebuttal{Although a larger database could be generated using implicit approaches based on, e.g., deep neural networks, it is unclear whether these techniques could achieve the same level of controllability to introduce highly specific geometric modifications, as we did here.}
\rebuttal{An effort was made to isolate the simultaneous effects of AF (heart rate reduction, wall kinetics, cardiac output) from the anatomic effects produced by adverse atrial remodeling. This effort included, in addition to performing fixed-wall simulations, considering the same heart rate, flow profile, and cardiac output range for AF and non-AF anatomies.}

Our results indicate that decreasing cardiac output \rebuttal{and increasing LAA volume accentuates LAA stasis, consistent with previous computational models \cite{Garcia-Isla2018, garcia2021demonstration, duenas2022morphing} and clinical data \cite{yamamoto2014complex,korhonen2015left,beinart2011left}. These trends are also consistent with the physical intuition that lower cardiac output implies slower flow velocities, whereas larger chamber volumes take longer to clear out. The dependence of LAA clearance on ostium is less trivial, as it involves two competing mechanisms. On one hand, increasing the ostium area while keeping the LAA emptying fraction constant likely slows down the LAA filling and emptying jets via mass conservation. On the other, secondary swirling motions caused by shear between fluid masses in the LAA and the LA body are likely to increase when their area of contact, i.e., the ostium area, increases \cite{drost2014parameterization, shankar2000fluid}. 
The first mechanism should dominate the flow in LAAs with normal or moderate emptying fractions (EF). However, secondary flows become dominant as the LAA EF approaches zero.
Clinically, a larger ostium area is associated with stasis \cite{yaghi2015left,lee2017additional} since it is usually associated with a larger volume of LAA and reduced cardiac output, making it difficult to evaluate the independent contribution of each parameter to LAA stasis. In our fixed-wall simulations, the flow inside the LAA is entirely formed by secondary swirling motions, and therefore, the LAA stasis decreases with the ostium area.}

\rebuttal{The multifactorial nature of LAA blood stasis has hindered the derivation of simple models for} clinical decision support. While it has long been recognized that anatomical factors such as LAA morphology \rebuttal{\cite{DiBiase2012, khurram2013relationship,yaghi2020left}} and atrial body sphericity \cite{bisbal2013left} affect the risk of ischemic stroke in patients with AF, there is a lack of systematic approaches to evaluate these effects. \rebuttal{Moreover, since secondary swirling motions become increasingly dominant with worsening LA function, atrial flow patterns in AF patients are more sensitive to changes in anatomical and hemodynamic parameters \cite{duran2023pulmonary, gonzalo2022non}.} The proposed simulation framework based on rigging and morphing can help assess each patient's susceptibility to LAA thrombosis due to atrial remodeling. Furthermore, we can make some generalizable observations by quantifying the effect of different variables independently. 
For example, with adverse atrial remodeling, one can expect both LAA volume and the ostium area to increase with time. However, while increasing the volume of LAA seems to increase LAA stasis indefinitely, the effect of the ostium area saturates rapidly, producing little benefit in lowering LAA stasis for areas larger than 3 cm$^2$.  \rebuttal{These results suggest that sufficiently adverse atrial remodeling should increase LAA stasis and, thus, a higher risk of thrombus formation.}

The second methodological innovation of this work is a framework for building reduced-order models of hemodynamic metrics in a universal LAA coordinate system (ULAAC).  \rebuttal{We applied singular value decomposition methods to derive reduced-order models that could be linked to atrial remodeling phenotypes and LAA blood stasis determinants. Previous studies have showcased deep learning models to infer LAA blood stasis from LAA geometry \cite{saiz2022unsupervised,pons2022joint}. While the boundaries between these two approaches are becoming increasingly blurry, singular value decomposition methods, such as POD, have the notable advantage of requiring less data to train and are computationally inexpensive.}

The ULAAC coordinates are defined by solving two Laplace-Beltrami equations on the LAA surface with different boundary conditions.   Roney et al. \cite{roney2019universal} previously used this method to define a universal whole-atrium coordinate system, but in that representation, the LAA typically occupies a small area that varies from patient to patient. By restricting our mapping to the LAA and prescribing fixed boundary conditions at the ostium and the LAA tip line, we map every patient's LAA into a unit square. This procedure helps to visualize and compare the spatial distribution of hemodynamic metrics in the LAA, the most likely site of thrombus formation. \rebuttal{This approach has the advantage of producing a bijective correspondence between each pair of coordinates ($\delta$, $\gamma$) on each point on the LAA surface, which may not be guaranteed in existing approaches \cite{acebes2021cartesian, morales2021deep}.}

To illustrate our methodology, we analyze TAWSS and OSI, two metrics associated with mean and temporal fluctuations of wall shear stress\rebuttal{, together with ECAP, the ratio between them.} TAWSS maps in the ULAAC system revealed a relatively common pattern of high TAWSS values near the ostium and much lower values in the distal LAA. This pattern was robust with respect to changes in cardiac output, LAA volume, and ostium area despite variations in fine-scale features. There is a trend for TAWSS to increase with cardiac output, consistent with the effect of this parameter on LAA stagnant volume. 
\rebuttal{Similarly, ECAP maps projected in the ULAAC system also revealed a common pattern of high values near the LAA tip and lower near the ostium. Consistent with the TAWSS maps, there is a trend for ECAP to decrease with cardiac output.}
However, it is harder to deduce the effects of the other two parameters and the presence of AF by mere visual inspection of patient-specific TAWSS \rebuttal{and ECAP} maps, and this problem is exacerbated in OSI maps, which have a more intricate spatial structure with higher patient-to-patient variability.

Our working hypothesis is that applying POD to the \rebuttal{hemodynamic} maps in the ULAAC coordinate system allows capturing the underlying patterns in these metrics and evaluating whether these patterns are distinct in LAAs with large stagnant volumes. We restricted our analysis to the first four POD modes because they represent more than $90\%$ of the variance of \rebuttal{TAWSS, OSI, and ECAP}. The POD eigenvectors describe the spatial organization of the \rebuttal{hemodynamic} fields, whereas the eigenvalues measure the weight of each mode.  
Regarding the TAWSS, these weights tend to cluster near zero for AF \rebuttal{anatomies}, especially those with large LAA stagnant volume, implying that the overall magnitude of TAWSS is lower throughout the appendage in these cases. The OSI eigenvalues show less clear segregation, although there is a palpable trend for mode 2 to have higher positive weights in \rebuttal{AF anatomies} with significant LAA stasis.  \rebuttal{The spatial organization of this mode shows positive values along the LAA tip line and negative values along the ostium. Lastly, for the ECAP eigenvalues, there is an appreciable trend for mode 1 to have higher weights in AF volumes with high stagnant volumes. Similarly to TAWSS eigenvalues, AF and non-AF anatomies tend to cluster together.} Overall, the TAWSS and OSI POD data imply that LAA blood stasis is more likely in appendages with low wall shear stresses, especially in those where stresses oscillate significantly near the LAA tip \rebuttal{\cite{Garcia-Isla2018, pons2022joint}}.

\subsection{Limitations and Future Work}

We simulate flow through rigid atria to reflect impaired wall kinetics associated with AF. 
\rebuttal{Fixed-wall CFD simulations of LA flow differ significantly from moving-wall simulations in normal atria and less substantially in atria with impaired function \cite{duenas2021comprehensive, garcia2021demonstration}. Thus, our non-AF simulations should represent normal LA anatomy but not normal LA function. 
Although this is a limitation, we believe this is an acceptable compromise for this study, since considering fixed walls allows the dissecting of geometrical and functional effects of atrial remodeling, providing information on the determinants of LAA stasis. We believe that the non-AF anatomy simulations provide valuable data by capturing the geometry of the atrium before adverse remodeling occurs.
In addition, running fixed-wall simulations considerably reduces the cost of segmenting and meshing  CT images, rigging, and morphing and allows CFD simulations to run faster. These simplifications allowed us to reach a total number of 180 simulations, which is significantly higher than most previous CFD studies of LA hemodynamics, allowing the evaluation of dimensionality reduction techniques.} 

\rebuttal{No MV model was considered in the atrium outflow, similar to most previous CFD investigations of the LAA flow. This choice is justified by simulations of the whole left heart \cite{Vedula2015}, suggesting that MV leaflet movement has little effect on LA flow patterns and, therefore, on LAA washing.}
\rebuttal{Similarly, the inflow for all simulations was based on a patient-specific AF profile, which limited the patient-specificity of the analysis but allowed us to isolate the effect of the flow profile from the atrial remodeling effects.} Furthermore, we imposed equal flow rates at the four PV atrial inlets. Although LAA stasis has been shown to vary with PV flow rate split, particularly in patients with AF \cite{duran2023pulmonary}, \rebuttal{the even flow split approximation is widely used when patient-specific data on this parameter are not available \cite{Lantz2018a}. Keeping the flow split and other patient-specific parameters constant throughout our simulations (wall kinetics, blood viscosity, heart rate, PV orientations, etc.) allowed us to isolate the effects of specific variables on LAA stasis.} \rebuttal{We also assumed blood to behave as a Newtonian fluid, even if non-Newtonian effects could be significant in LAAs with low shear stress and stasis \cite{gonzalo2022non, zhang2023computational}}. 

\rebuttal{Regarding the generative methodology based on anatomical rigging and morphing, future efforts will address user dependence by fully automating the definition of the \textit{bones}, which may be achieved via skeletonization. Regarding the proposed ULAAC method, we are currently working to extend the parameterization to 3D/4D parameters, such as blood velocities and vortex dynamics, that could be relevant for characterizing LAA stasis behavior. Extending to time can be done by performing spatiotemporal POD \cite{borja2023deriving} while 
3D POD is trivial once a 3D ULAAC system is available.}

Despite leveraging a relatively large database of CFD models, the baseline patient-specific segmentations for these models come from only 6 subjects \rebuttal{varying only one parameter at a time and constituting only a small sample among all possible LAA/LA morphologies}. Our results show that the 30 simulations of each patient tend to cluster together in the POD eigenvalue space, especially for TAWSS \rebuttal{and ECAP}, suggesting that these simulations are not independent. Consequently, this work should be viewed as a first effort toward applying modal decomposition to hemodynamic metrics to identify and interpret the determinants of LAA thrombosis, rather than as an exhaustive demonstration of this idea. More atrial geometries are needed to confirm and generalize the results obtained here and to fully characterize how atrial remodeling affects LAA stasis\rebuttal{, with the exciting possibility to study additional parameters such as the LAA bending angle \cite{yaghi2020left}}. Despite these limitations, the notion that hemodynamic metrics of each patient leave a distinct low-dimensional footprint that withstands variations in geometrical parameters and cardiac output is compelling and deserves further investigation. \rebuttal{Moreover, combined with the flexibility of our rigging and morphing methods, reduced-order modeling of hemodynamic metrics could shed light onto the key morphological parameters determining LAA stasis. }


\section*{Acknowledgments} 
This work was supported by \textit{Ministerio de Ciencia, Innovación y Universidades of Spain} under projects DPI 2017-83911-R and PID 2019-107279RB-I00, by \textit{Junta de Castilla y León} under project ``Proyecto de Apoyo a GIR 2018'' with reference VA081G18, by \textit{Junta de Extremadura} and FEDER funds under project IB20105, and by the US National Institutes of Health under projects 1R01HL160024 and 1R01HL158667. We thank the \textit{Programa Propio - Universidad Politécnica de Madrid}, and the \textit{Ayuda Primeros Proyectos de Investigación ETSII-UPM}. We also thank \textit{Programa de Excelencia para el Profesorado Universitario de la Comunidad de Madrid} for its financial support and the CeSViMa UPM project for its computational resources.

\section*{Conflict of interest}
The authors declare that they have no conflict of interest.






\bibliographystyle{elsarticle-num} 
\bibliography{cas-refs}




\end{document}